\documentstyle[12pt,epsfig,subfigure,color,soul,fleqn, rotating]{article}

\newcommand{\be}{\begin{eqnarray}}
\newcommand{\ee}{\end{eqnarray}}
\newcommand{\nee}{\nonumber\end{eqnarray}}

\newcommand{\mch}[1] {m_{\ti \x^+_{#1}}}
\newcommand{\mnt}[1] {m_{\ti \x^0_{#1}}}

\newcommand{\msg}    {m_{\ti g}}

\def\fb              {{\rm fb}^{-1}}
\def\gev             {{\rm GeV}}

\newcommand{\gsim}{\;\raisebox{-0.9ex}
           {$\textstyle\stackrel{\textstyle >}{\sim}$}\;}
\newcommand{\lsim}{\;\raisebox{-0.9ex}{$\textstyle\stackrel{\textstyle<}
           {\sim}$}\;}

\def\a               {\alpha}
\def\b               {\beta}
\def\d               {\delta}

\def\x               {\chi}
\def\ti              {\tilde}
\def\sq              {\ti q}
\def\st              {\ti t}

\def\ch              {\ti \x^\pm}

\def\nt              {\ti \x^0}
\def\sg              {\ti g}

\newcommand{\AddrVienna}{
\it Universit\"at Wien, Fakult\"at f\"ur Physik,
A-1090 Vienna, Austria \\}

\newcommand{\AddrGAKUGEI}{%
 \it Department of Physics, Tokyo Gakugei University, Koganei,
Tokyo 184-8501, Japan\\}

\newcommand{\AddrHEPHY}{%
 \it Institut f\"ur Hochenergiephysik der \"Osterreichischen Akademie
der Wissenschaften, A-1050 Vienna, Austria\\}

\newcommand{\AddrWuerzburg}{%
 \it Institut f\"ur Theoretische Physik und Astrophysik, Universit\"at W\"urzburg,
D-97074 W\"urzburg, Germany\\}

\newcommand{\AddrDESY}{
\it  Deutsches Elektronen-Synchrotron (DESY), Theory Group,
D-22603 Hamburg, Germany \\}

\begin{document}

\begin{flushright}
DESY 10-059
\\UWThPh-2011-22
\end{flushright}

\bigskip
\bigskip

\begin{center}
  \textbf{\LARGE
Flavour violating gluino three-body decays at LHC}\\[10mm]
\large{A. Bartl${}^{1}$, H. Eberl${}^2$, E. Ginina${}^{1}$, B. Herrmann${}^{3}$, K.~Hidaka${}^4$,\\
W.~Majerotto${}^2$ and W. Porod${}^{5}$
}
\vspace{0.5cm}\\
\small{
$^1$ \AddrVienna
$^2$ \AddrHEPHY
$^3$ \AddrDESY
$^4$ \AddrGAKUGEI
$^5$ \AddrWuerzburg}
\end{center}

\noindent

%
\thispagestyle{empty}

\begin{abstract}
We study the effect of squark generation mixing on gluino production and
decays at LHC in the Minimal Supersymmetric Standard Model (MSSM) for the case
that the gluino is lighter than all squarks and dominantly decays into three particles, $\sg \to q \bar{q} \nt_k,~q \bar{q}' \ch_l$.
We assume mixing between the second and the third squark generations in the up-type and down-type
squark sectors.
We show that this mixing can lead to very large branching ratios of the quark-flavour
violating gluino three-body decays despite the strong constraints on
quark-flavour violation (QFV) from the experimental data on B mesons. We also show that the
QFV gluino decay branching ratios are very sensitive not only to the
generation mixing in the squark sector, but also to the parameters of the neutralino and chargino
sectors.
We show that the branching ratio of the QFV
gluino decay $\sg \to c \bar{t} (\bar{c} t) \nt_1$ can go up to
$\sim 40\%$. Analogously,
that of the QFV decay $\sg \to s \bar{b} (\bar{s} b) \nt_1$ can reach $\sim 35 \%$.
We find that the rates of the resulting QFV signatures, such as $pp \to t t \bar{c} \bar{c} E_{\rm T}^{mis}$, can be
significant at LHC. This could have an important influence on the gluino searches at LHC.

\end{abstract}
\clearpage


\section{Introduction}
\label{sec:intro}

The flavour structure of the quark sector is very well described by the
Cabibbo-Kobayashi-Maskawa (CKM) mixing matrix, which is the only source of quark flavour
violation (QFV) in the Standard Model (SM). In particular, flavour changing neutral
current (FCNC) processes, such as $K^0 \to \mu^+ \mu^-$, $B^0 \to \mu^+ \mu^-$,
$B \to X_s \gamma$, $B \to X_s\,l^+ l^-$ etc., are strongly suppressed~\cite{PDG2008}.
They impose strong constraints on the quark generation mixing. Any
extension of the SM must therefore respect these constraints.

In supersymmetric (SUSY) extensions of the SM, mixing between different quark flavours
in the squark sector, that is not related to the CKM-matrix is also possible. Although the
mixing between the first and second generation squarks is strongly constrained,
there is room for appreciable mixing between the second and third generation of squarks,
still obeying the constraints from B meson data\footnote{There could also be mixing between the right up-squark and the left top-squark  which is hardly constrained~\cite{Plehn}. However, we do not consider this mixing here.}. This is beyond the minimal flavour violation
(MFV), where the only source of  QFV is the mixing due to the CKM matrix~\cite{Buras:2000dm,Ambrosio:2002ex,Kagan:2009bn}.

The effects of mixing between the second and third squark generations, especially the mixing between top and
charm squarks, have been studied in the Minimal Supersymmetric Standard Model (MSSM) for
squark production and decays at LHC~\cite{K1,K2,Ba1,Bj, Hurth:2009ke}. Squark generation mixing has been investigated
in detail in the QFV decays of gluinos, $\ti{g} \to \bar{\ti{u}}_{1,2}\,c ~ (\ti{u}_{1,2}\,\bar{c}) \to c\, \bar{t}\, \nt_1 \,(\bar{c}\, t \,\nt_1)$,
where $\ti{u}_{1,2}$ are the lightest squark states and are mixtures of charm and top squarks~\cite{Ba2}. There it is assumed
that at least one squark is lighter than the gluino, so that the gluino decays first into a real squark (antisquark) and an antiquark (quark). In~\cite{Ba2} it is shown that
this leads to pronounced edge structures in the charm top-quark invariant mass distribution.

In the present paper we study the QFV gluino decays in the general MSSM assuming that all squarks are heavier than the gluino,
so that the gluino dominantly
decays into three particles. This will give rise to a very different pattern of QFV gluino decays
as compared to the studies in Ref.~\cite{Ba2},
due to interference effects between the various virtual squark exchange contributions.
Moreover, the
invariant mass distributions will have different shapes without any edge structure in contrast to the case studied in~\cite{Ba2}.

We study the mixing between the second and the third generations not only in the up-type squark sector as in Ref.~\cite{Ba2},
but also in the down-type sector. We investigate QFV gluino decays including those into down-type quark pair plus
neutralino, such as $\sg \to s \bar{b} \nt_1$. Furthermore,
we also investigate in detail the dependence of the QFV gluino decay branching ratios on the neutralino/chargino parameters. We
take into account all relevant experimental constraints on the MSSM parameters from B
physics and searches for Higgs bosons and SUSY particles, and the theoretical constraints on the trilinear couplings from the vacuum stability conditions.

Recently ATLAS and CMS performed searches for SUSY at LHC with $\sqrt{s}=7$ TeV on the basis of total 
integrated luminosities of $35~\rm{pb}^{-1}$ \cite{ATLAS, CMS} and $\sim$~1~fb$^{-1}$ \cite{new_ATLAS, new_CMS}. 
They found no excess of events over the SM expectations and set limits on the squark and gluino 
masses at $95\%$ confidence level (CL).
In the simplified SUSY model with $m_{\tilde{\chi}^0_1} \approx 0$ (where all SUSY particles other than the gluino and  
squarks of the first two generations, and $\tilde{\chi}^0_1$ are decoupled by being given very heavy masses) 
the limit on the gluino mass is $m_{\tilde{g}} \gsim 800$ GeV for large $m_{\tilde{q}}$ and that on the squark mass 
is $m_{\tilde{q}} \gsim 850$ GeV for large $m_{\tilde{g}}$ \cite{new_ATLAS} (see \cite{new_CMS} also). Here, $m_{\tilde{q}}$
is the degenerate mass of the squarks of the first two generations.
In the context of the constrained MSSM (CMSSM) (or mSUGRA) the lower limit on the gluino mass 
is smaller than $\sim$650 GeV for $m_{\tilde{q}} \gsim 1.5$ TeV and the limit on the squark mass is $m_{\tilde{q}} \gsim 1.1$ TeV 
for any $m_{\tilde{g}}$ \cite{new_CMS, new_ATLAS_1} (see \cite{new_ATLAS} also). Here again $m_{\tilde{q}}$ is defined 
to be the degenerate mass of the squarks of the first two generations and the masses of the third 
generation squarks are (significantly) smaller than this $m_{\tilde{q}}$ in this framework of the CMSSM (mSUGRA). 
Therefore we will assume a gluino mass of about 1 TeV in our analysis respecting these 
gluino mass limits.

\section{Squark mixing with flavour violation}
\label{sec:sq.matrix}

In the MSSM the most general form of the squark mass matrices in the super-CKM basis of $\sq_{0 \gamma} =
(\sq_{1 {\rm L}}, \sq_{2 {\rm L}}, \sq_{3 {\rm L}},
\sq_{1 {\rm R}}, \sq_{2 {\rm R}}, \sq_{3 {\rm R}}),~\gamma = 1,...6,$
where $(q_1, q_2, q_3) = (u, c, t), (d, s, b)$ is~\cite{Allanach:2008qq}
\begin{equation}
    {\cal M}^2_{\tilde{q}} = \left( \begin{array}{cc}
        {\cal M}^2_{\tilde{q},LL} & {\cal M}^2_{\tilde{q},LR} \\[2mm]
        {\cal M}^2_{\tilde{q},RL} & {\cal M}^2_{\tilde{q},RR} \end{array} \right),
    \label{EqMassMatrix}
\end{equation}
for $\tilde{q}=\tilde{u},\tilde{d}$, where the $3\times3$ matrices read
\begin{eqnarray}
    {\cal M}^2_{\tilde{d},LL} = M_Q^2 + D_{\tilde{d},LL}{\bf 1} + \hat{m}^2_d, &\qquad&
    {\cal M}^2_{\tilde{u},LL} = V_{\rm CKM} M_Q^2 V_{\rm CKM}^{\dag} + D_{\tilde{u},LL}{\bf 1} + \hat{m}^2_u, \nonumber \\
    {\cal M}^2_{\tilde{d},RR} = M_D^2 + D_{\tilde{d},RR}{\bf 1} + \hat{m}^2_d, & &
    {\cal M}^2_{\tilde{u},RR} = M_U^2 + D_{\tilde{u},RR}{\bf 1} + \hat{m}^2_u.
    \label{EqM2LLRR}
\end{eqnarray}
Here $M^2_{Q,U,D}$ are the hermitian soft-SUSY-breaking mass matrices of the squarks and
$\hat{m}_{u,d}$ are the diagonal mass matrices of up- and down-type quarks.
$D_{\tilde{q},LL} = \cos 2\beta m_Z^2 (T_3^q-e_q
\sin^2\theta_W)$ and $D_{\tilde{q},RR} = e_q \sin^2\theta_W \cos 2\beta m_Z^2$,
where
$T_3^q$ and $e_q$ are the isospin and
electric charge of the quarks (squarks), respectively, and $\theta_W$ is the weak mixing
angle.
The left-left blocks of the up-type and down-type squark mass matrices are related
by the CKM matrix $V_{\rm CKM}$ due to the SU$(2)_{\rm L}$ symmetry. Note that $V_{\rm CKM} M_Q^2 V_{\rm CKM}^{\dag} \simeq M_Q^2$
as $V_{\rm CKM} \simeq 1$.
The off-diagonal blocks of eq.\
(\ref{EqMassMatrix}) read
\begin{eqnarray}
    {\cal M}^2_{\tilde{d},RL} = {\cal M}^{2\dag}_{\tilde{d},LR} &=& \frac{v_1}{\sqrt{2}} T^T_D - \mu^* \hat{m}_d\tan\beta, \nonumber \\
    {\cal M}^2_{\tilde{u},RL} = {\cal M}^{2\dag}_{\tilde{u},LR} &=& \frac{v_2}{\sqrt{2}} T^T_U - \mu^* \hat{m}_u\cot\beta \,.
\end{eqnarray}
Here $T^T_{U,D}$ are transposes of $T_{U,D}$ which are the soft-SUSY-breaking trilinear coupling matrices of the up-, down-type  squarks: ${\cal L}_{int} \supset -(T_{U\alpha \beta} \ti{u}^\dagger _{R\beta}\ti{u}_{L\alpha}H^0_2+T_{D\alpha \beta} \ti{d}^\dagger _{R\beta}\ti{d}_{L\alpha}H^0_1)$,
$\mu$ is the higgsino mass parameter, and $\tan\beta=v_2/v_1$, where $v_{1,2}=\sqrt{2} \left\langle H^0_{1,2} \right\rangle$
are the vacuum expectation values of the neutral Higgs fields.
The squark mass matrices are diagonalized by the $6\times6$ unitary matrices $R^{\tilde{q}}$,
$\tilde{q}=\tilde{u},\tilde{d}$, such that
\begin{equation}
    R^{\tilde{q}} {\cal M}^2_{\tilde{q}} (R^{\tilde{q} })^{\dag} = {\rm diag}(m_{\tilde{q}_1}^2,\dots,m_{\tilde{q}_6}^2)
    \qquad{\rm with}\qquad m_{\tilde{q}_1} < \dots < m_{\tilde{q}_6} \,\,.
\end{equation}
The physical mass eigenstates $\sq_i, i=1,...,6$ are given by $\sq_i =  R^{\sq}_{i \a} \sq_{0\a} $.

In accordance with~\cite{Gabbiani1996} we define the QFV parameters
$\delta^{LL}_{\alpha\beta}$, $\delta^{uRR}_{\alpha\beta}$
and $\delta^{uRL}_{\alpha\beta}$ $(\alpha \neq \beta)$ as follows:
\begin{eqnarray}
\delta^{LL}_{\alpha\beta} & \equiv & M^2_{Q \alpha\beta} / \sqrt{M^2_{Q \alpha\alpha} M^2_{Q \beta\beta}}~,
\label{eq:InsLL}\\[3mm]
\delta^{uRR}_{\alpha\beta} &\equiv& M^2_{U \alpha\beta} / \sqrt{M^2_{U \alpha\alpha} M^2_{U \beta\beta}}~,
\label{eq:InsRR}\\[3mm]
\delta^{uRL}_{\alpha\beta} &\equiv& (v_2/\sqrt{2} ) T_{U\beta\alpha} / \sqrt{M^2_{U \alpha\alpha} M^2_{Q \beta\beta}}~.
\label{eq:InsRL}
\end{eqnarray}
%
\begin{table}
\small{
\caption{Constraints on the MSSM parameters from the B-physics experiments relevant mainly for the mixing between
the second and the third generations of squarks, and from the Higgs sector.
The fourth column shows constraints at $95 \%$ CL
obtained by combining the experimental error quadratically with the theoretical uncertainty, except for $B(B_s \to \mu^+ \mu^-)$
and $m_{h^0}$, which is the mass of the lighter CP-even neutral Higgs boson.
$R^{\rm{SUSY}}_{B \to \tau \nu}$$ \equiv $${\rm {B} (B^+ \to \tau^+ \nu)_{\rm {SUSY}}}\over{\rm {B}(B^+ \to \tau^+ \nu)_{\rm{SM}}}$
$\approx
[1-((m_{B^+} \tan \beta)/ m_{H^+})^2]^2$, where $ m_{H^+}$ is the charged Higgs boson mass~\cite{R00}.}
\begin{center}
\begin{tabular}{|c|c|c|c|}
    \hline
    Observable & Exp.\ data & Theor.\ uncertainty & \ Constr.\ (95$\%$CL) \\
    \hline\hline
    $\Delta M_{B_s}$ [ps$^{-1}$] & $17.77 \pm 0.12$ (68$\%$ CL)~\cite{PDG2008} & $\pm 3.3$ (95$\%$ CL)~\cite{R22} & $17.77 \pm 3.31$\\
    $10^4\times$B($b \to s \gamma)$ & $3.55 \pm 0.26$ (68$\%$ CL)~\cite{Tr} & $\pm 0.23$ (68$\%$ CL)~\cite{BsgTheo2} &  $3.55\pm 0.68$\\
    $10^6\times$B($b \to s~l^+ l^-$)&&&\\
    $(l=e~{\rm or}~\mu)$ & $1.60 \pm 0.50$ (68$\%$ CL)~\cite{R5} & $\pm 0.11$ (68$\%$ CL)~\cite{Bsll} & $1.60 \pm 1.00$\\
    $10^8\times$B($B_s\to \mu^+\mu^-$) & $ < 4.3$  (95$\%$ CL)~\cite{btomumu} && $<4.3$ \\
    $10^4\times$B($B^+ \to \tau^+ \nu $) & $1.68 \pm 0.31$  (68$\%$ CL)~\cite{Tr} & $\pm0.25$  (68$\%$ CL)~\cite{Tr} & $R^{\rm{SUSY}}_{B \to \tau \nu}=$\\&&&$1.40 \pm 0.76$\\
    $m_{h^0}$ [GeV] & $ > 114.4$ (95$\%$ CL)~\cite{PDG2008, R8} & $\pm 4.0$~\cite{Allanach:2004rh} & $> 110.4$ \\
    \hline
\end{tabular}
\end{center}
\begin{center}
\end{center}
\label{TabConstraints}
\vspace*{-1cm}}
\end{table}
Here $\alpha,\beta=1,2,3 ~(\alpha \ne \beta)$ denote the quark flavours.
The QFV parameters in the up-type squark sector relevant for this study are $\delta^{LL}_{23}$, $\delta^{uRR}_{23}$,
$\delta^{uRL}_{23} =( \delta^{uLR}_{32})^*$ and $ \delta^{uLR}_{23} = ( \delta^{uRL}_{32})^*$ which are the $\ti{c}_L - \ti{t}_L$, $\ti{c}_R-\ti{t}_R$,
$\ti{c}_R - \ti{t}_L$ and $\ti{c}_L - \ti{t}_R$ mixing parameters, respectively.
For the down-type squark sector we define the QFV parameters as follows ($\a \ne \b$):
\begin{eqnarray}
\delta^{dRR}_{\alpha\beta} &\equiv& M^2_{D \alpha\beta} / \sqrt{M^2_{D \alpha\alpha} M^2_{D \beta\beta}}~,
\\[3mm]
\delta^{dRL}_{\alpha\beta} &\equiv& (v_1/\sqrt{2} ) T_{D \beta\alpha} / \sqrt{M^2_{D \alpha\alpha} M^2_{Q \beta\beta}}~.
\end{eqnarray}
The QFV parameters in the down-type squark sector relevant for our study are
$\delta^{LL}_{23}$, $\delta^{dRR}_{23}$,
$\delta^{dRL}_{23} =\big( \delta^{dLR}_{32}\big)^*$ and $\delta^{dLR}_{23} = \big( \delta^{dRL}_{32}\big)^*$ which are the $\ti{s}_L - \ti{b}_L$, $\ti{s}_R-\ti{b}_R$,
$\ti{s}_R - \ti{b}_L$ and $\ti{s}_L - \ti{b}_R$ mixing parameters, respectively.

In our analysis we neglect mixing between the first two squark generations due to the severe experimental
constraints from K meson physics. We also neglect mixing between the first and third squark generations
focusing on the effects of mixing between the second and third generations.
We assume all the QFV parameters to be real. These parameters are also subject to the experimental constraints given in Table~\ref{TabConstraints}.

Furthermore we impose the vacuum stability
conditions for the trilinear coupling matrices~\cite{Casas}
\begin{eqnarray}
|T_{U\alpha\alpha}|^2 &<&
3~Y^2_{U\alpha}~(M^2_{Q \alpha\alpha}+M^2_{U\alpha\alpha}+m^2_2)~,
\label{eq:CCBfcU}\\[2mm]
|T_{D\alpha\alpha}|^2 &<&
3~Y^2_{D\alpha}~(M^2_{Q\alpha\alpha}+M^2_{D\alpha\alpha}+m^2_1)~,
\label{eq:CCBfcD}\\[2mm]
|T_{U\alpha\beta}|^2 &<&
Y^2_{U\gamma}~(M^2_{Q \alpha\alpha}+M^2_{U\beta\beta}+m^2_2)~,
\label{eq:CCBfvU}\\[2mm]
|T_{D\alpha\beta}|^2 &<&
Y^2_{D\gamma}~(M^2_{Q\alpha\alpha}+M^2_{D\beta\beta}+m^2_1)~,
\label{eq:CCBfvD}
\end{eqnarray}
where
$\a,\b=1,2,3,~\a\neq\b;~\gamma={\rm Max}(\a,\b)$ and
$m^2_1=(m^2_{H^\pm}+m^2_Z\sin^2\theta_W)\sin^2\beta-\frac{1}{2}m_Z^2$,
$m^2_2=(m^2_{H^\pm}+m^2_Z\sin^2\theta_W)\cos^2\beta-\frac{1}{2}m_Z^2$.
The Yukawa couplings of the up-type and down-type quarks are
$Y_{U\alpha}=\sqrt{2}m_{u_\alpha}/v_2=\frac{g}{\sqrt{2}}\frac{m_{u_\alpha}}{m_W \sin\beta}$
$(u_\a=u,c,t)$ and
$Y_{D\alpha}=\sqrt{2}m_{d_\alpha}/v_1=\frac{g}{\sqrt{2}}\frac{m_{d_\alpha}}{m_W \cos\beta}$
$(d_\a=d,s,b)$,
with $m_{u_\a}$ and $m_{d_\a}$ being the running quark masses at the weak scale and
$g$ being the SU(2) gauge coupling. All soft-SUSY-breaking parameters are also assumed to be given
at the weak scale. As SM parameters we take $m_W=80.4~\gev$, $m_Z=91.2~\gev$ and
the on-shell top-quark mass $m_t=173.3~\gev$ \cite{Shabalina}.
We have found that our results shown in the following are fairly insensitive to the
precise value of $m_t$.

In addition to the constraints in Table~\ref{TabConstraints} we impose the following limits:
\begin{enumerate}
    \item The limit on ($m_{A^0}$, $\tan\beta$) from the negative search for neutral MSSM Higgs bosons
          decaying into a tau pair (i.e.\ $A^0/H^0/h^0 \rightarrow \tau^+ \tau^-$) at LHC \cite{MSSM_Higgs_LHC}
          with $m_{A^0}$ being the mass of the CP-odd neutral Higgs boson $A^0$.                          
    \item The experimental limit on SUSY contributions on the electroweak $\rho$ parameter~\cite{R1}: 
          $\Delta \rho~ (\rm SUSY) < 0.0012.$
    \item The LEP limits on the SUSY particle masses~\cite{R2}: $m_{\ch_1} > 103~\gev, 
          m_{\nt_1} > 50~\gev$, where $m_{\ch_1}$ and $m_{\nt_1}$ are the masses of the 
          lighter chargino and the lightest neutralino, respectively.
\end{enumerate}

We take into account all the constraints listed in this section in all plots presented in this article.
The constraints on the QFV parameters from $B(b \to s \gamma), \Delta M_{B_s}$ and the vacuum stability conditions are especially important
for this study.
For the computation of the observables (i.e.\ physical masses, decay branching ratios,
$\Delta M_{B_s}$ and $\Delta \rho$(SUSY)) we use the public code SPheno  v3.0~\cite{R1:SPheno, Porod:2011nf} .

\section{QFV three-body decays of gluino}
\label{sec:gl.decays}

\begin{figure}
\begin{center}
\begin{tabular}{cc}
\resizebox{5.0cm}{!}{\includegraphics{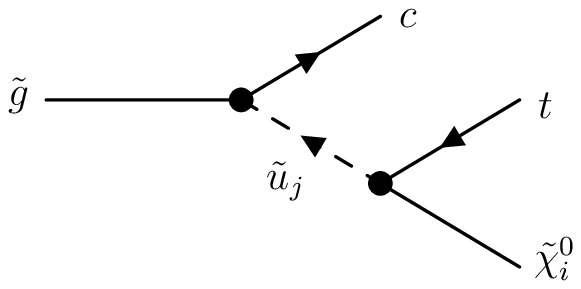}} & \hspace{1.cm}
\resizebox{5.0cm}{!}{\includegraphics{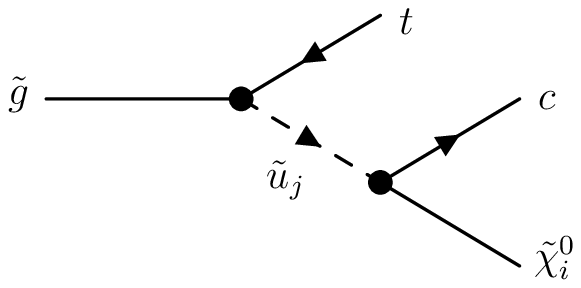}}
\end{tabular}
\end{center}
\caption{Feynman diagrams for $\sg \to c\, \bar{t}\, \nt_i$.}
\label{Feyngraphs}
\end{figure}
If all squarks are heavier than the gluino and squark generation mixing occurs only between
the second and third generation, one has the following QFV three-particle decays of gluino into quarks and neutralinos $\nt_i$, $i=1,2,3,4$,
\begin{eqnarray}
\ti{g} \to c\, \bar{t}\, \nt_i,\, \bar{c}\, t \,\nt_i\,,
\label{eq:decay1}\\
\ti{g} \to s\, \bar{b}\, \nt_i,\, \bar{s}\, b \,\nt_i\,.
\label{eq:decay2}
\end{eqnarray}
%
\begin{table}
\small{
\caption{Weak scale parameters at $Q=1~{\rm TeV}$ for our prototype QFV scenario, except for  $m_{A^0}$ which is the pole mass
(i.e.\ physical mass ) of $A^0$.
All of $T_{U \a \a}$ and $T_{D \a \a}$ are 0.}
\begin{center}
\begin{tabular}{|c|c|c|c|c|c|}
  \hline
 $M_1$ & $M_2$ & $M_3$ & $\mu$ & $\tan \beta$ & $m_{A^0}$ \\
 \hline \hline
 139~\gev  &  264~\gev &  800~\gev &  1000~\gev & 10 &  800~\gev \\
  \hline
\end{tabular}
\vskip0.4cm
\begin{tabular}{|c|c|c|c|}
  \hline
   & $\a=\beta = 1$ & $\a=\beta = 2$ & $\a=\beta = 3$ \\
  \hline \hline
   $M_{Q \a \beta}^2$ & $(3150)^2~\gev^2$ &  $(3100)^2~\gev^2$  & $(3050)^2~\gev^2$ \\
   \hline
   $M_{U \a \beta}^2$ & $(3000)^2~\gev^2$ & $(2200)^2~\gev^2$ & $(2150)^2~\gev^2$ \\
   \hline
   $M_{D \a \beta}^2$ & $(3000)^2~\gev^2$ & $(2990)^2~\gev^2$ &  $(2980)^2~\gev^2$  \\
   \hline
\end{tabular}
\end{center}
\label{msg972soft}}
\end{table}
\begin{table}[h!]
\small{
\caption{Physical masses of the particles in the scenario of Table~\ref{msg972soft}. $m_{H^0}$ is the mass of the heavier CP-even neutral Higgs
boson $H^0$.}
\begin{center}
\begin{tabular}{|c|c|c|c|c|c|}
  \hline
  $\mnt{1}$ & $\mnt{2}$ & $\mnt{3}$ & $\mnt{4}$ & $\mch{1}$ & $\mch{2}$ \\
  \hline \hline
  $139~\gev$ & $281.3~\gev$ & $1017.9~\gev$ & $1021.7~\gev$ & $281.5~\gev$ & $1022.7~\gev$ \\
  \hline
\end{tabular}
\vskip 0.4cm
\begin{tabular}{|c|c|c|c|c|}
  \hline
  $\msg$ & $m_{h^0}$ & $m_{H^0}$ & $m_{A^0}$ & $m_{H^+}$ \\
  \hline \hline
  $975~\gev$ & $121.1~\gev$  & $800.3~\gev$ & $800~\gev$ & $804~\gev$ \\
  \hline
\end{tabular}
\end{center}
\label{msg972masses}}
\end{table}

We will mainly focus on the decays into $\nt_1$.
The corresponding Feynman diagrams for the decay $\sg \to c \bar{t} \nt_i$
are shown in Fig.~\ref{Feyngraphs}.
We have interference between the $t$ and the $u$ channel exchanges, as well as between
the different $\ti{u}_j$ exchange diagrams. In particular, there can be a strong
destructive interference between the $\ti{u}_l$ and $\ti{u}_k$ contributions, if they
are mainly $\ti{c}$ and $\ti{t}$ mixtures and their masses are similar.
For instance, if $\ti{u}_l \sim \cos \theta \st_R+\sin \theta \ti{c}_R$ and
$\ti{u}_k \sim -\sin \theta \st_R+\cos \theta \ti{c}_R$
then the $\ti{u}_l$ exchange contribution is $\sim \frac{(+\cos \theta \sin \theta) }{(p^2 -m_{\ti{u}_l}^2)}$ whereas the
$\ti{u}_k $ exchange contribution is $\sim \frac{(-\cos \theta \sin \theta) }{(p^2 -m_{\ti{u}_k}^2)}$.
These two contributions almost cancel with each other for $m_{\ti{u}_l}\approx m_{\ti{u}_k}$. The suppression of this cancellation requires a large
mass-splitting between the two squarks which can be induced by a large $\ti{c}_R - \ti{t}_R$ mixing term $M^2_{U_{23}}$
(or $\delta^{uRR}_{23}$) even in case the $\ti{c}_R$ mass parameter $M^2_{U_{22}}$ is similar to the $\ti{t}_R$ mass parameter  $M^2_{U_{33}}$.
Moreover, in this case one has a very strong $\ti{c}_R - \ti{t}_R$ mixing. Therefore one can expect sizable QFV decay branching ratios
for a large mass-splitting and hence for large values of $\delta^{uRR}_{23}$.
For the decays into down-type quarks one has analogous Feynman
diagrams with the replacements $c \to s,\, t \to b, \,\ti{u}_j \to
\ti{d}_j $.
\begin{figure}
\centering
\subfigure[]{
   { \mbox{\hspace*{-0.6cm} \resizebox{8.4cm}{!}{\includegraphics{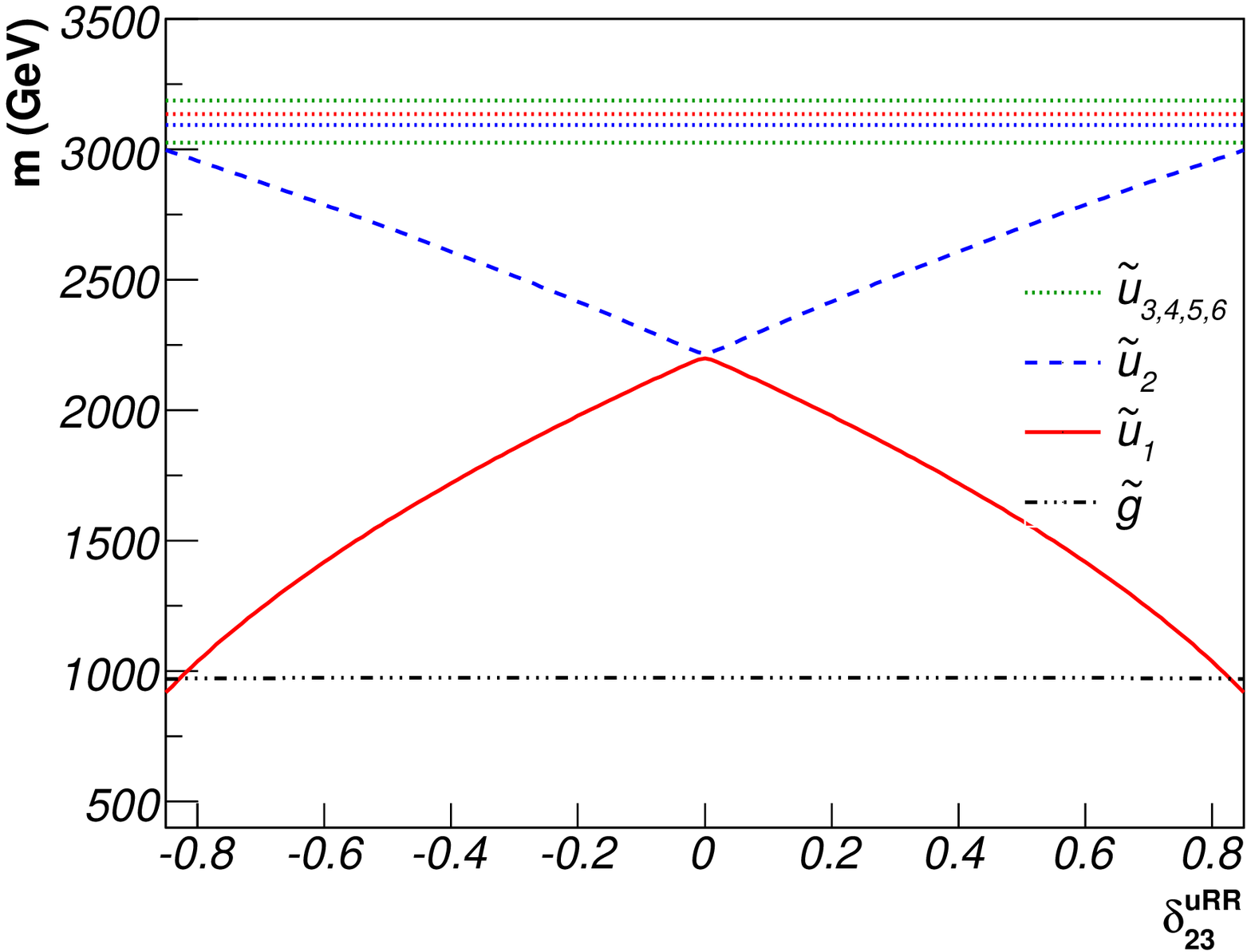}}\hspace*{-0.8cm} }}
   \label{A13masses}}
 \subfigure[]{
   { \mbox{\hspace*{-0.3cm} \resizebox{8.4cm}{!}{\includegraphics{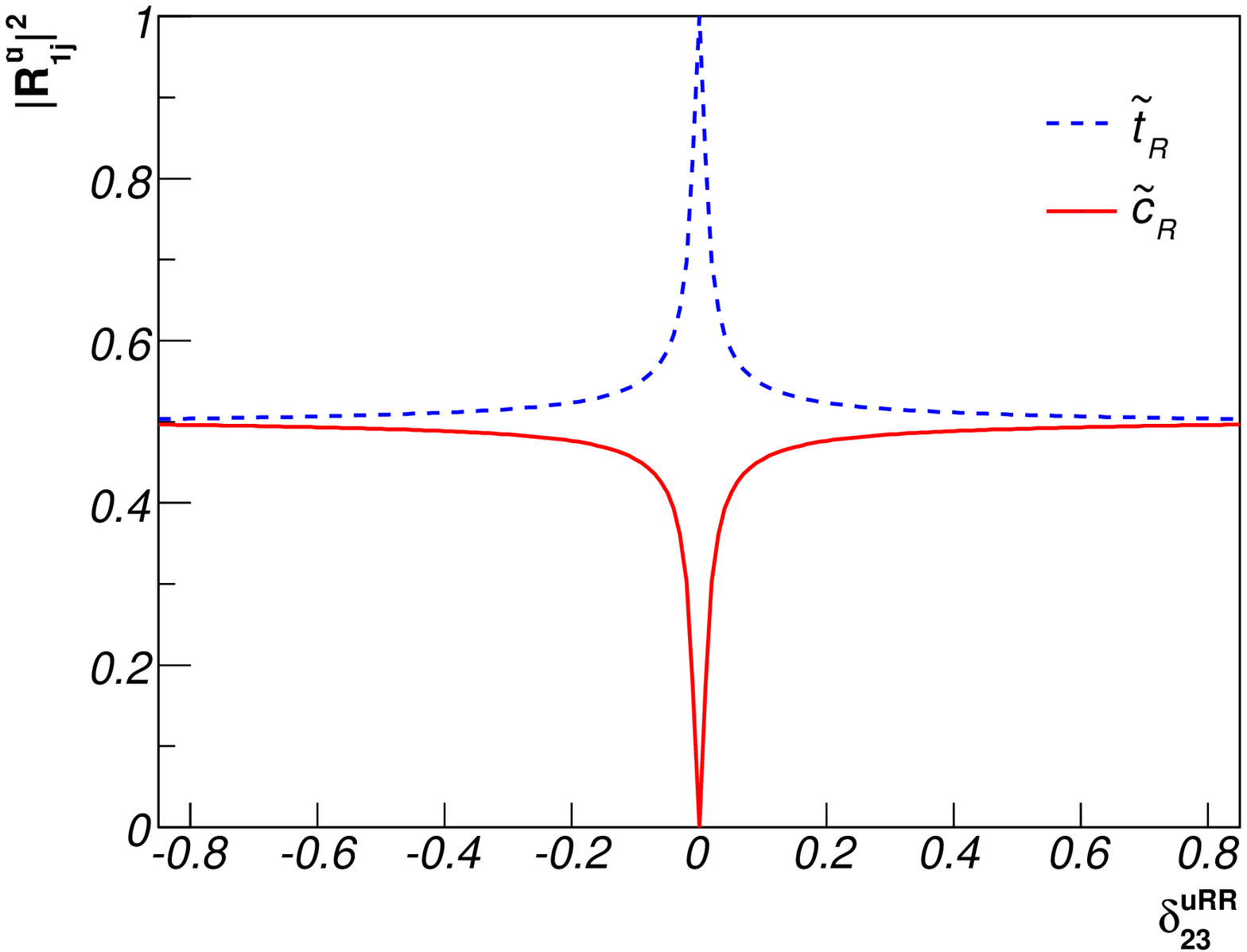}}\hspace*{-1cm}}}
  \label{A13su1decomp}}
\caption{(a) Up-type squark and gluino masses and (b) flavour decomposition of $\ti{u}_1$ (i.e.\ $|R^{\ti{u}}_{15}|^2 \equiv |\ti{c}_R~ {\rm component} |^2 $(full red line) and   $|R^{\ti{u}}_{16}|^2 \equiv |\ti{t}_R~ {\rm component} |^2  $ (dashed blue line))
as functions of the QFV parameter
$\delta^{uRR}_{23}$, with the other QFV parameters being zero, for the scenario of Table~\ref{msg972soft}. \label{A13}}
\end{figure}

There are also gluino three-body decays into charginos, such as
$\sg \to c \bar{b} \ti{\ }^-_1,~ s \bar{t} \ti{\chi}^+_1$, etc..
We will, however, not discuss them explicitly here, although they
are included in our branching ratio calculations.

We calculate the three-particle decay branching ratios of the gluino according to the diagrams in Fig.~\ref{Feyngraphs} and their charge conjugated ones, including the QFV couplings given in~\cite{K1}.
As basic SUSY parameters at the weak
scale we take
$ M_1, M_2, M_3, \mu, \tan \beta, m_{A^0},
M^2_{Q \a \b},$ $ M^2_{U \a \b}, M^2_{D \a \b}, T_{U \a \b}$ and $T_{D \a \b}$,
which we assume to be real. Here $M_{1,2,3}$ are the U(1), SU(2) and SU(3) gaugino mass parameters, respectively, and $m_{A^0}$ is the pole mass
(i.e.\ physical mass) of the Higgs boson $A^0$.
We study in detail the QFV scenario based on the parameters of Table~\ref{msg972soft},
given at  the scale $Q=1$ TeV,
according to the SPA convention~\cite{SPA} except for $m_{A^0}$ being the pole mass of $A^0$. The scenario of Table~\ref{msg972soft}
satisfies all the constraints listed in Section~\ref{sec:sq.matrix}; e.g.\ for the low energy observables we obtain
$\Delta M_{B_s}=19.01~{\rm ps}^{-1}, ~{\rm B}(b \to s \gamma)=3.46\times10^{-4}, ~ {\rm B}(b \to s\,l^+ l^-)=1.59\times 10^{-6}, ~
 {\rm B}(b \to s\, \mu^+ \mu^-)=5.02\times 10^{-9}, ~R^{SUSY}_{B \to \tau \nu} =0.99, ~ m_{h^0}=121.1~\gev, ~ \Delta \rho~({\rm SUSY}) =5.70\times 10^{-5}$.
We add to the parameters of Table~\ref{msg972soft} the QFV
parameters $\delta^{LL}_{23}, \delta^{uRR}_{23}, \delta^{uRL}_{23}, \delta^{uLR}_{23}$ as well as $\delta^{dRR}_{23}, \delta^{dRL}_{23}, \delta^{dLR}_{23}$ (given also at  $Q=1$ TeV), and vary them in a range allowed by the constraints listed in Section~\ref{sec:sq.matrix}.
The physical masses for the case with all the QFV parameters being zero are shown in Table~\ref{msg972masses}.
They are calculated from the basic MSSM parameters at the one-loop level, taking into account
the complete flavour structure~\cite{R1:SPheno}.
We have found that these masses are fairly insensitive
to the QFV parameters.

Note that in our case QFV left-right mixing effects, i.e.\ those due to $\delta^{uRL}_{23},\delta^{uLR}_{23}, $ $\delta^{dRL}_{23}, \delta^{dLR}_{23}$
cannot be significant. We show this for the left-right mixing parameter $\delta^{uRL}_{23}$.
Due to the vacuum stability condition (\ref{eq:CCBfvU}) we have
$|T_{U32}|^2 \lsim Y^2_{U3} (M^2_{Q33}+M^2_{U22}+m_2^2) \approx M^2_{Q33}+M^2_{U22} \approx $ O(10 TeV$^2$), because $Y_{U3} \approx 1$ and $m_2^2 \ll M^2_{Q33}+M^2_{U22}$. Therefore, $|\delta^{uRL}_{23}|= \frac{v_2}{\sqrt{2}}\frac{|T_{U32}|}{\sqrt{M^2_{Q33} M^2_{U22}}}\lsim
\frac{v_2}{\sqrt{2}} \sqrt{\frac{M^2_{Q33}+ M^2_{U22}}{M^2_{Q33} M^2_{U22}}}\approx 0.1$.
Analogously, the parameters $\delta^{uLR}_{23},\delta^{dRL}_{23},\delta^{dLR}_{23}$ are also constrained to be very small due to the vacuum stability conditions.
Therefore, the most relevant QFV parameters in our study are $\delta^{LL}_{23}, \delta^{uRR}_{23}, \delta^{dRR}_{23}$.

Fig.~\ref{A13}a shows the physical masses of the up-type squarks $\ti{u}_1,...,\ti{u}_6$ as functions of the QFV parameter $\delta^{uRR}_{23}$,
with all the other QFV parameters being zero, for the scenario of Table~\ref{msg972soft}.
All the constraints mentioned in Section~\ref{sec:sq.matrix} are fulfilled in the shown range.
Masses of all the down-type squarks $\ti{d}_i$ are about 3 TeV in this range.
For $|\delta^{uRR}_{23}| \lsim 0.8$ all squarks are heavier than the gluino.
In Fig.~\ref{A13}b we show the flavour decomposition of $\ti{u}_1$. For $|\delta^{uRR}_{23}| \gsim 0.2$, $\ti{u}_1$ is practically a full mixture of
$\ti{c}_R$ and $\ti{t}_R$. For $|\delta^{uRR}_{23}| \lsim 0.88$ the flavour decomposition of $\ti{u}_2$ is similar to that of $\ti{u}_1$ with $\ti{c}_R$ and
$\ti{t}_R$ interchanged.
\begin{figure}
\vspace*{0.5cm}
{\setlength{\unitlength}{1mm}
\begin{center}
\begin{picture}(135,50)
\put(25,0){ \mbox{\resizebox{7.5cm}{!}{\includegraphics{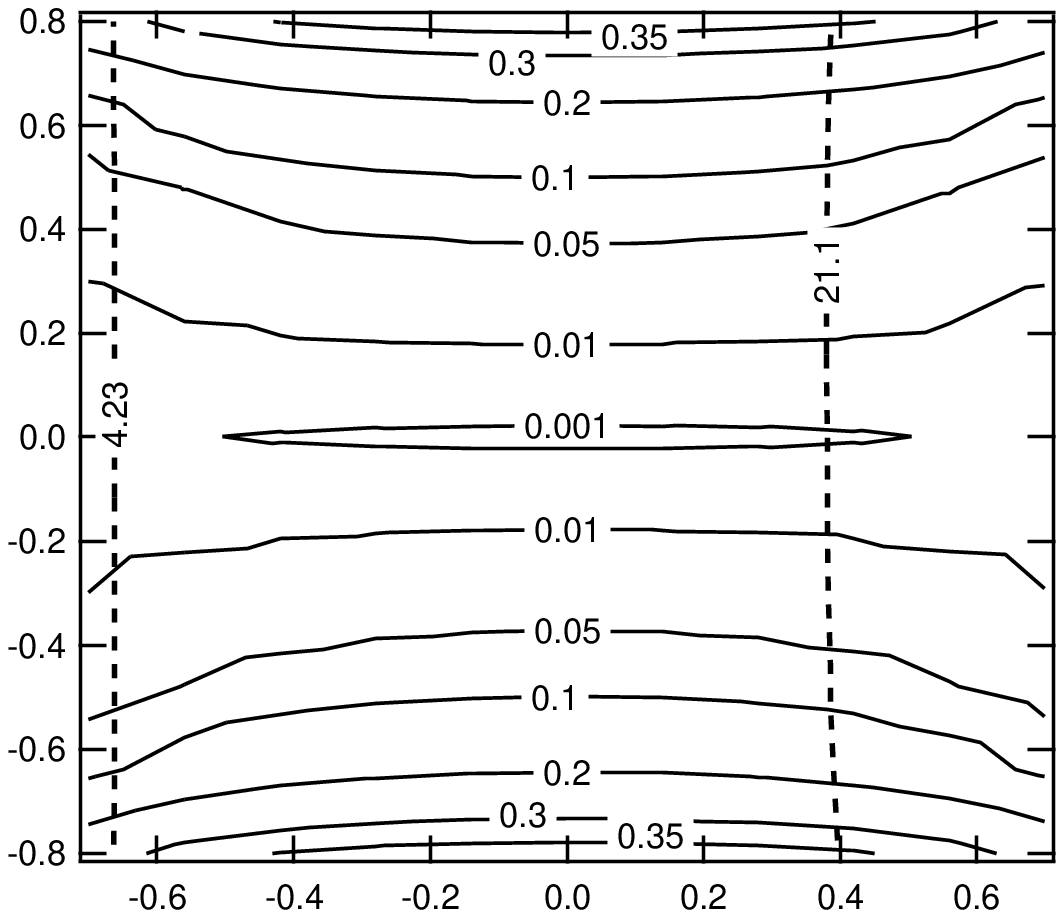}}}}
\put(61,-2){ \mbox{ $\delta^{LL}_{23}$}}
\hspace*{2.2cm}
\begin{sideways}
\put(29,17){\mbox{ $\delta^{uRR}_{23}$}}
\end{sideways}
\end{picture}
\end{center}}
\caption{Contours of the QFV decay branching ratio B($\sg \to c t \nt_1$)
in the $\delta^{LL}_{23}$ - $\delta^{uRR}_{23}$ plane with the other QFV parameters being zero, for the scenario of Table~\ref{msg972soft} (solid lines).
Also shown are the contours
of $\Delta M_{B_s} = 21.1~{\rm ps}^{-1}$ and  $10^{4}\times $B$(b \to s \gamma) = 4.23$ (dashed lines). The region between the two dashed lines
is allowed by all the constraints mentioned in Section~\ref{sec:sq.matrix}, including those from $\Delta M_{B_s}$  and B$(b \to s \gamma)$. } \label{BRctcontour}
\end{figure}

In Fig.~\ref{BRctcontour} we show contours of the branching ratio
${\rm B}(\sg \to c t \nt_1) \equiv {\rm B}(\sg \to c \bar{t} \nt_1) + {\rm B}(\sg \to \bar{c} t \nt_1)$ in the $\delta^{LL}_{23} - \delta^{uRR}_{23}$
plane together with contours of $10^{4}\times$B$(b \to s \gamma) = 4.23$ and $\Delta M_{B_s} = 21.1~\rm{ps}^{-1}$, where the other
parameters are fixed as in Table~\ref{msg972soft}. The branching ratio B($\sg \to c t \nt_1 $) can reach values larger than 35\%. 
As can be seen, in the range $-0.65 \lsim
\delta^{LL}_{23} \lsim 0.35$, the B$(b \to s \gamma)$ and the $\Delta M_{B_s}$ constraints are fulfilled. 
In Ref.~\cite{Heinemeyer:2004by} it is shown that also the $\rho$-parameter data
may constrain the flavour off-diagonal elements of the squark mass matrices, in particular
the $\delta^{LL}_{\alpha \beta}$ entries. However, in our case the $\rho$-parameter
practically does not give constraints for two reasons. First, we have negligible left-right 
squark mixing implying that the left-squark sector is almost decoupled from the 
right-squark sector. Second, the mass matrices of the left up-type squarks and the left 
down-type squarks are related by the $SU(2)_L$ symmetry leading to approximately the same 
masses and mixing matrices for the up-type and down-type squarks.

\begin{figure}
\begin{center}
\resizebox{8.8cm}{!}{\includegraphics{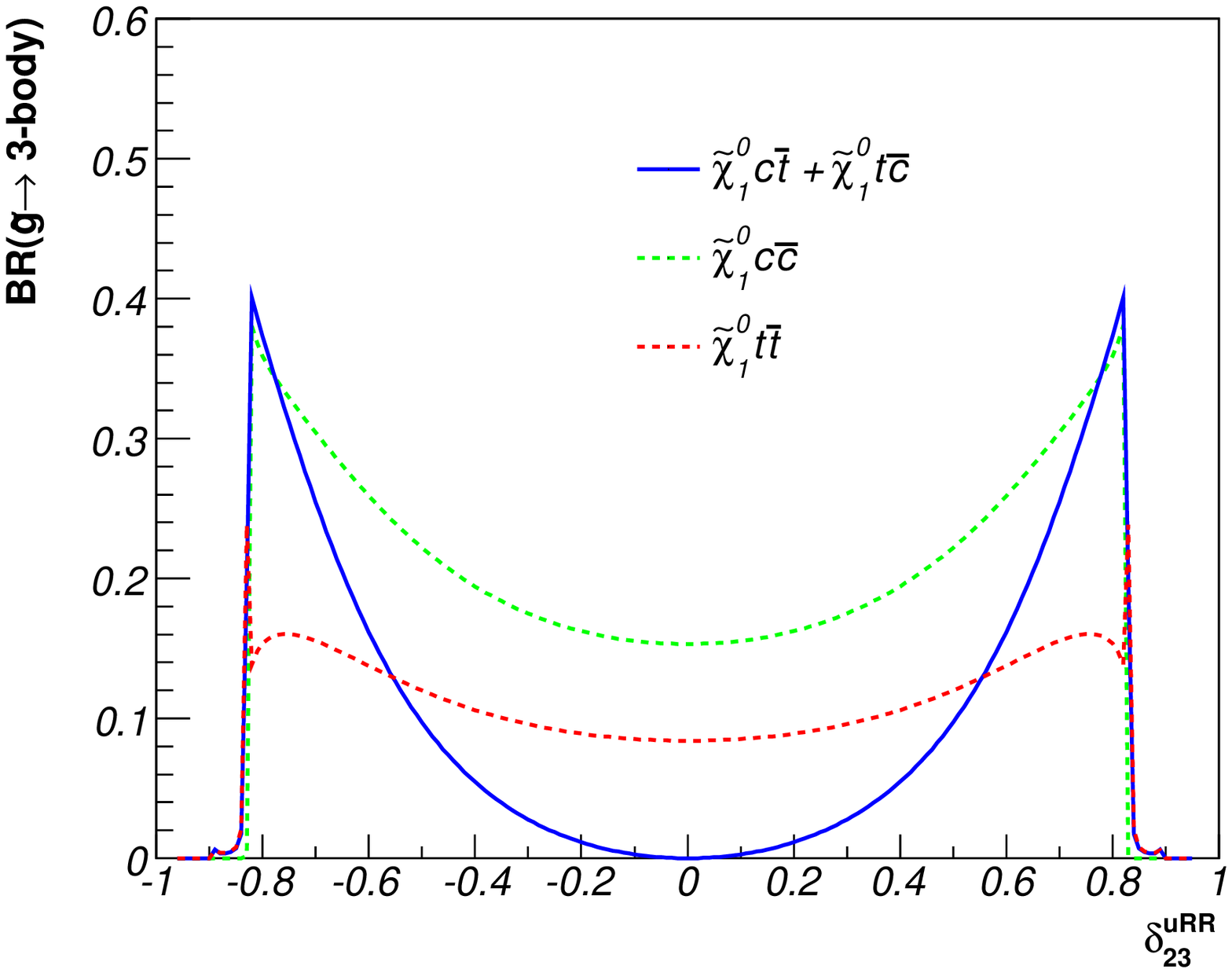}}
\end{center}
\caption{The branching ratios of the decays $\sg \to c \bar{t} \nt_1 + \bar{c} t \nt_1$, $\sg \to c \bar{c} \nt_1$ and $\sg \to t \bar{t} \nt_1$ as functions of $\d^{uRR}_{23}$ for the other QFV parameters being zero and the other parameters are fixed as in Table~\ref{msg972soft}.
} \label{gBRs}
\end{figure}

In Fig.~\ref{gBRs} the branching ratios B($\sg \to c t \nt_1$), B($\sg \to c \bar{c} \nt_1$) and B($\sg \to t \bar{t} \nt_1$) are shown as functions of $\d^{uRR}_{23}$, with the other QFV parameters being zero and the other parameters fixed as in Table~\ref{msg972soft}. All the constraints mention in Section~\ref{sec:sq.matrix} are fulfilled in the shown range. One can see that
the QFV decay branching ratio B($\sg \to  c t \nt_1$) can reach
40\% and
that in the range $0.6 \lsim |\delta^{uRR}_{23}| \lsim 0.8$ the QFV decay branching ratio B($\sg \to c t \nt_1$) is even larger than the quark-flavour conserving (QFC) branching ratio B($\sg \to t \bar{t} \nt_1 )$.
For $|\delta^{uRR}_{23}| \gsim 0.8$, the two-body decays into $\ti{u}_1$ dominate because $\ti{u}_1$ becomes lighter than the gluino
(see Fig.~\ref{A13}a). The reason for
this large QFV decay
branching ratio is as follows: For $0.6 \lsim |\d^{uRR}_{23}| \lsim 0.8$, all squarks
other than $\ti{u}_1$ (including down-type squarks) are very heavy (see Fig.~\ref{A13masses}), which leads to the dominance of the
$\ti{u}_1$ exchange contribution in the gluino decays. In this $\d^{uRR}_{23}$ range the $\ti{u}_1, ~\ti{u}_2$ are strong
mixtures of $\ti{c}_R$
and $\ti{t}_R$ and the mass-splitting between $\ti{u}_1$ and $\ti{u}_2$ is very large,
preventing a strong
destructive interference between the $\ti{u}_1$ and $\ti{u}_2$ exchange contributions
in this $\d^{uRR}_{23}$ range
(see Fig.~\ref{A13}). This gives the large QFV decay branching ratio B$(\sg \to c
t \nt_1)$. 
Note
that $\ti{u}_1 ( \sim \ti{c}_R + \ti{t}_R)$ couples to $\nt_1 ( \approx \ti{B}^0)$ and practically
does not couple to
$\nt_2 ( \approx \ti{W}^0)$ and $\ch_1 ( \approx \ti{W}^\pm)$ (see Table~\ref{msg972soft}). Moreover, $\nt_{3,4}$ and $\ch_2$ are very heavy
in the QFV scenario considered here (see Table~\ref{msg972masses}.).
$\ti{B}^0$ and $\ti{W}^{0, \pm}$
are the U(1) and SU(2) gauginos (the bino and winos), respectively.
\begin{figure}[h!]
\vspace*{0.5cm}
{\setlength{\unitlength}{1mm}
\begin{center}
\begin{picture}(135,50)
\put(25,0){ \mbox{\resizebox{8.1cm}{!}{\includegraphics{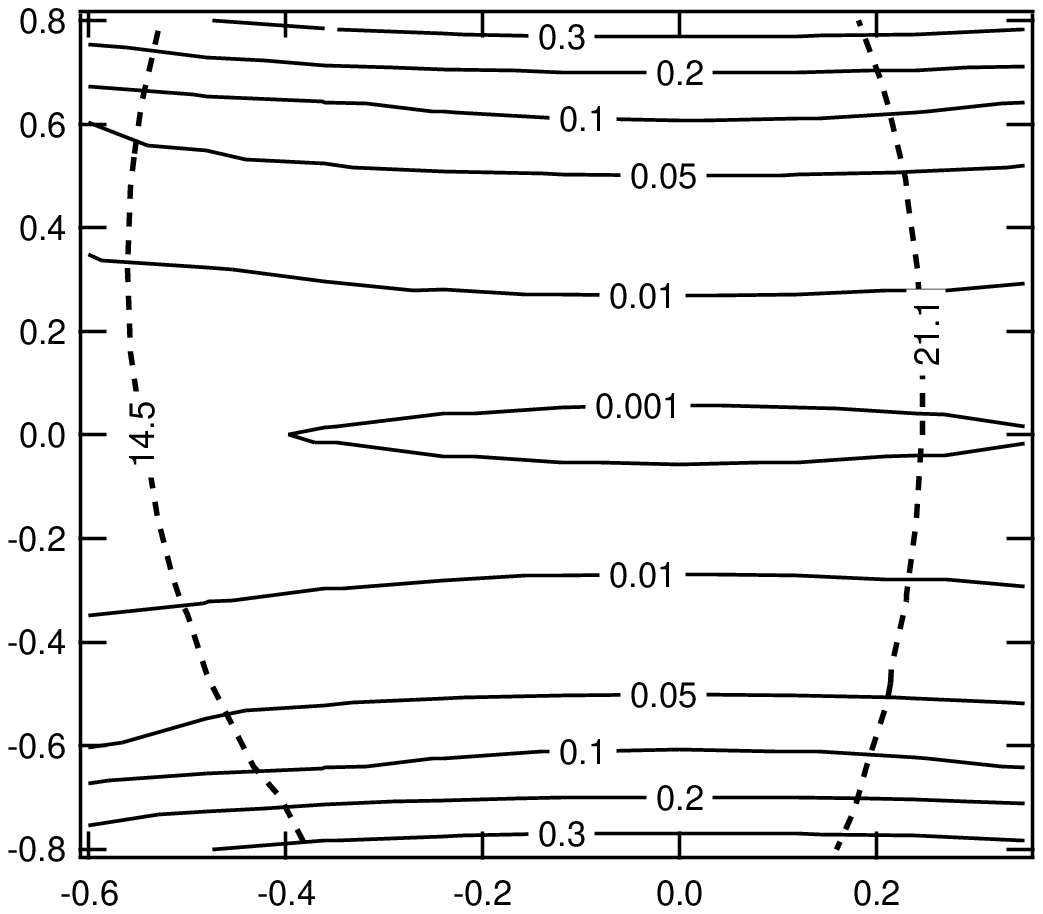}}}}
\put(64,-2){ \mbox{ $\delta^{LL}_{23}$}}
\hspace*{2.3cm}
\begin{sideways}
\put(32,17){\mbox{ $\delta^{dRR}_{23}$}}
\end{sideways}
\end{picture}
\end{center}}
\caption{Contours of the QFV decay branching ratio B($\sg \to s b \nt_1$) in the $\delta^{LL}_{23}$ - $\delta^{dRR}_{23}$ plane, with the other QFV parameters
being zero for the scenario of Table~\ref{msg972soft}, but with the values of $M^2_{U \a \a}$ and $M^2_{D \a \a}$ interchanged  (solid lines).
Also shown are the contour
lines for $\Delta M_{B_s} = 14.5~{\rm ps}^{-1}$ and $\Delta M_{B_s} = 21.1~{\rm ps}^{-1}$ (dashed lines). The region between the two dashed lines
is allowed by all the constraints mentioned in Section~\ref{sec:sq.matrix}, including the $\Delta M_{B_s}$ constraint. } \label{BRsbcontour}
\end{figure}

On the other hand, as can be seen in  Fig.~\ref{BRctcontour}, the dependence of B$(\sg \to c t \nt_1)$ on the $\ti{c}_L$ - $\ti{t}_L$ mixing parameter $\delta^{LL}_{23}$
is much weaker than that on $\delta^{uRR}_{23}$.
This is mainly due to the fact that in our scenario the left-squarks
$(\ti{c}_L, \ti{t}_L)$ are significantly heavier than the right-squarks $(\ti{c}_R, \ti{t}_R)$
and that the left-squark coupling to $\nt_1$ ($\sim$ bino) is small.
Hence, the contributions of the left-squark exchanges to
$B(\sg \to c t \nt_1)$ are suppressed, leading to the very small effect of the  $\ti{c}_L - \ti{t}_L$
mixing parameter $\delta^{LL}_{23}$ on the QFV decay branching ratio.
As a consequence, for $|\delta^{uRR}_{23}| \lsim 0.4$ the QFV decay branching ratio B($\sg \to c t \nt_1$) is smaller than $5\%$ even for larger allowed values of $|\delta^{LL}_{23}|$.

The gluino can also have QFV decays into down-type quarks with sizeable branching ratios if the $\delta^{dRR}_{23}$ or $\delta^{LL}_{23}$ are unequal to zero. As an example, in Fig.~\ref{BRsbcontour} we show a contour plot of
the QFV decay branching ratio $ {\rm B}(\sg \to s b \nt_1) \equiv {\rm B}(\sg \to s \bar{b} \nt_1)+ {\rm B}(\sg \to \bar{s} b \nt_1)$ in the $\delta^{dRR}_{23}$ - $\delta^{LL}_{23}$ plane for the scenario of Table~\ref{msg972soft}.
The dashed contour lines for $\Delta M_{B_s} = 14.5~\rm{ps}^{-1}$ and $21.1~\rm{ps}^{-1}$ show
the region $-0.38 \lsim
\delta^{LL}_{23} \lsim 0.12$
allowed by all the
constraints mentioned in Section~\ref{sec:sq.matrix}, including the B$(b \to s \gamma)$ constraint.
Note that in this case the $\Delta M_{B_s}$ constraint is significantly stronger than the
B$(b \to s \gamma)$ constraint in the whole $\delta^{LL}_{23} - \delta^{dRR}_{23}$ plane.
$ {\rm B}(\sg \to s b \nt_1)$ can reach values larger than 30$\%$. The reason for this sizable QFV decay branching ratio is similar to that for the large QFV decay branching ratio B$(\sg \to c t \nt_1)$.
The dependence of this
QFV decay branching ratio on $\delta^{LL}_{23}$ is again much weaker than that on $\delta^{dRR}_{23}$.

We also would like to note that we have found a scenario giving
QFV three-body decay branching ratios B($\sg \to c t \nt_1$) (or
B($\sg \to s b \nt_1$)) of about 50\% for a gluino mass of $\sim 1$ TeV, still satisfying all the
relevant constraints. In such a scenario, however, the heavier
squarks have masses of about 6 TeV, while the lightest
squark heavier than the gluino has a mass of about 1 TeV.

\section{Influence of the neutralino/chargino parameters on the QFV three-body gluino decays}
\label{sec:neucha}

\begin{figure}
\vspace*{0.5cm}
{\setlength{\unitlength}{1mm}
\begin{center}
\hspace*{-20mm}
\begin{picture}(125,60)
\put(25,0){ \mbox{
\resizebox{8.1cm}{!}{\includegraphics{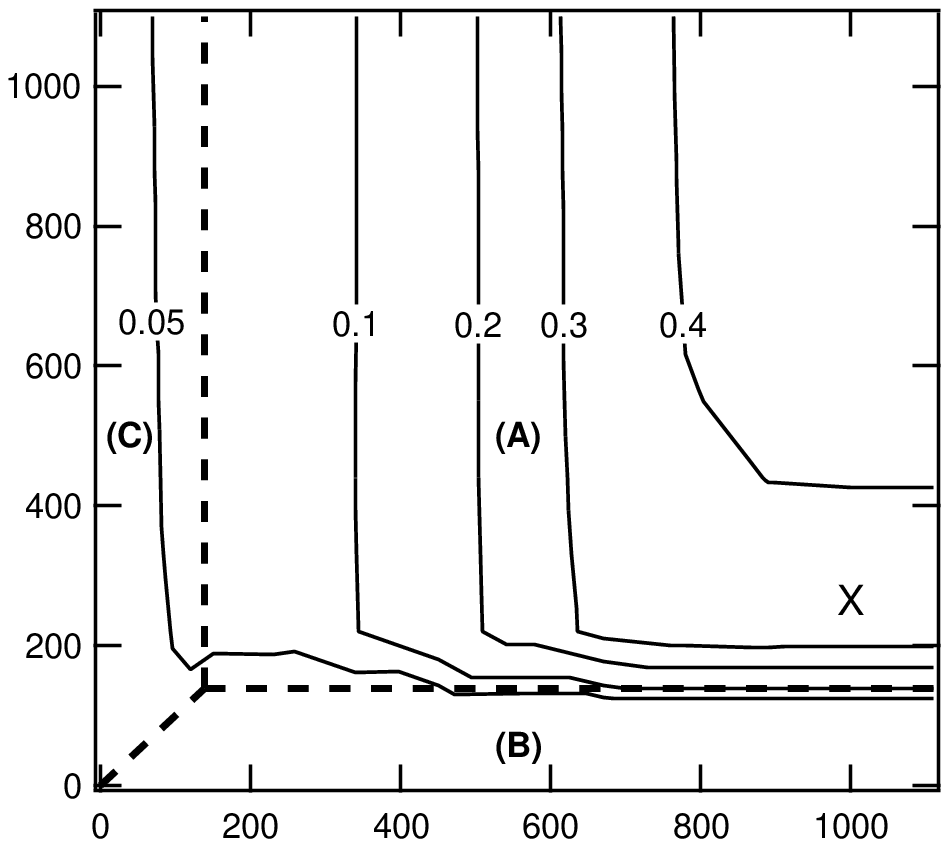}}
}}
\put(65,-1){ \mbox{ $\mu~(\gev)$}}
\hspace*{2.7cm}
\begin{sideways}
\put(27,12){\mbox{$M_2~(\gev)$}}
\end{sideways}
\end{picture}
\end{center}}
\caption{Contour plot for $B(\sg \to c t \nt_{1})$ (solid lines) in the $\mu - M_2$ plane for
$\delta_{23}^{uRR} = 0.8$, the other QFV parameters being zero,
and the other parameters specified as in Table~\ref{msg972soft} with $M_1=139~\gev$.
Region (A): bino-like LSP region; region (B):
wino-like LSP
region; region (C): higgsino-like LSP region.
The point "X" corresponds to our reference scenario given in Table~\ref{msg972soft}:
$M_2= 264~\gev,~\mu= 1000~\gev.$
}
\label{muM2dependence}
\end{figure}

As the squark generation mixing enters into the
squark-quark-neutralino/chargino couplings,
here we study how the pattern of the QFV gluino decays depends on the
parameters of the
neutralino-chargino sector. First we show in Fig.~\ref{muM2dependence}  a contour plot of the
branching ratio
$B(\sg \to c t \nt_{1})$ in the $\mu - M_2$ plane for $\delta_{23}^{uRR} = 0.8$, the other QFV parameters
being zero, and the other parameters fixed as in Table~\ref{msg972soft}. In the
whole plane $m_{\sg} \approx 972~\gev$.
As one can see,
this branching ratio is larger than $10 \%$ for $\mu \gsim 350~\gev$. We indicate
the regions where the
LSP is bino-, wino-, or higgsino-like. The largest QFV decay branching ratio
$B(\sg \to c t \nt_{1})$ is in
the bino-like LSP region reaching up to $\sim 40\%$.
%

\begin{figure}
\centering
\subfigure[]{
   { \mbox{\hspace*{-0.6cm} \resizebox{8.4cm}{!}{\includegraphics{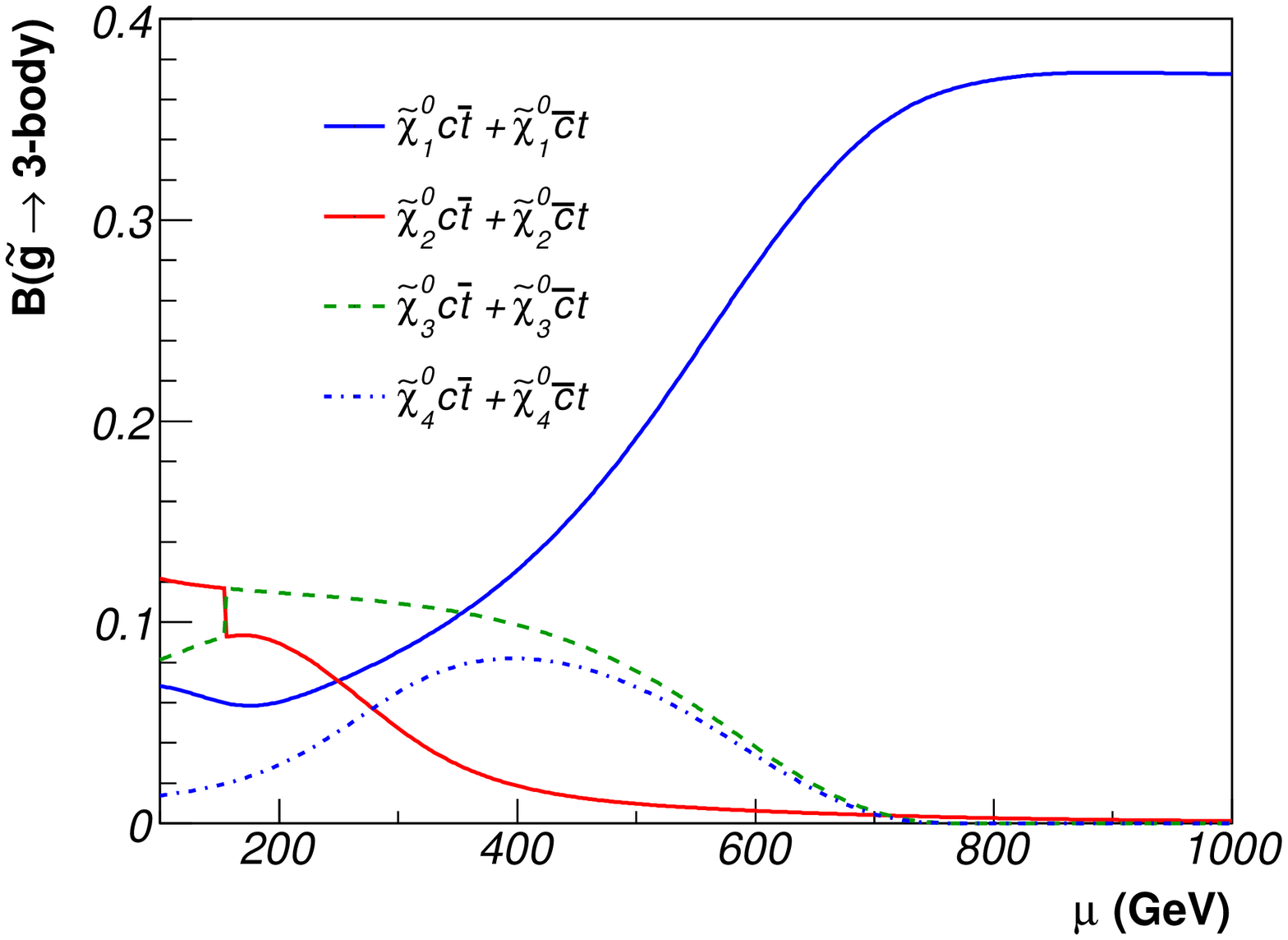}} \hspace*{-0.8cm}}}
   \label{mudepend1}}
 \subfigure[]{
   { \mbox{\hspace*{-0.3cm} \resizebox{8.4cm}{!}{\includegraphics{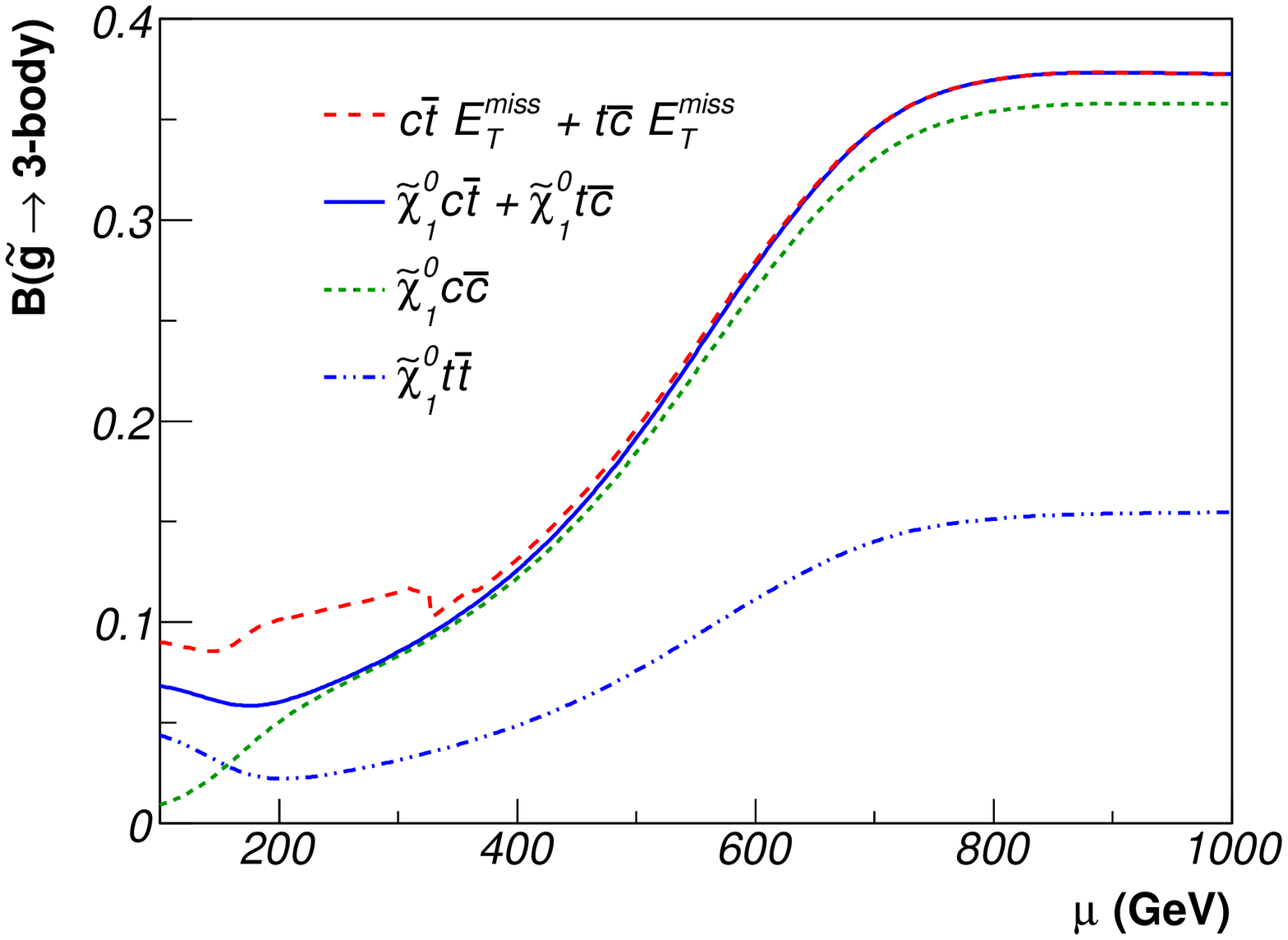}} \hspace*{-1cm}}}
  \label{mudepend2}}
\caption{The $\mu$ dependence of the QFV and QFC gluino decay branching ratios for
$\delta^{uRR}_{23}=0.8$, the other QFV parameters being zero, for the scenario given in Table~\ref{msg972soft}.
(a) Branching ratios of the decays $\sg \to c \bar{t} \nt_i + \bar{c} t \nt_i$, $i=1,...,4$ as a function of $\mu$.
(b) Branching ratios of the decays $\sg \to c \bar{c} \nt_1$, $\sg \to t \bar{t} \nt_1$,
 $\sg \to c \bar{t} \nt_1 + \bar{c} t \nt_1$ and $\sg \to c \bar{t} E_{\rm T}^{mis}+ \bar{c} t E_{\rm T}^{mis}$ as a function of $\mu$.
 \label{BRsM1139}}
\end{figure}

In the following
we discuss the dependence of the QFV decay branching ratios of the gluino on the higgsino mass parameter $\mu$ in more detail. In Fig.~\ref{BRsM1139} we show this dependence
for $\delta^{uRR}_{23}=0.8$, the other QFV parameters being zero, and the other parameters fixed as in Table~\ref{msg972soft}. 
In the whole range $m_{\sg} \approx 972~\gev$ and the constraints mentioned in Section~\ref{sec:sq.matrix} are satisfied.
The branching ratios of the QFV gluino decays $\sg \to c \bar{t} \nt_i+\bar{c} t \nt_i, ~i=1,...,4$ are shown in Fig.~\ref{BRsM1139}a.
For $\mu \gsim M_1$ the bino component of $\nt_1$ is increasing and hence the branching ratio of $\sg \to c \bar{t} (\bar{c} t) \nt_1$ increases.
For $\mu \gsim 700~\gev$ this branching ratio is about 36$\%$, being roughly a factor of 2 larger than that for  $\sg \to t \bar{t} \nt_1$.
For $|\mu| \lsim M_1$, $\nt_1$ and $\nt_2$ are essentially higgsinos, and the branching ratio of $\sg \to c \bar{t} (\bar{c} t) \nt_1$ is less than 10$\%$. The
reason is that the QFV decays into the higher neutralinos and the charginos become more important.
Note that $|\mu|\lsim 100~\gev$ is excluded by the LEP limit on the $m_{\ch_1}$.
Furthermore, the final state $c \bar{t} E_{\rm T}^{mis}+\bar{c} t E_{\rm T}^{mis}$
can contain contributions from the higher neutralino modes $c \bar{t} (\bar{c} t)
\nt_i$,
$i \geq 2$, with the invisible decays of $\nt_i \to \nt_1 \nu \bar{\nu}$, where $E_{\rm T}^{mis}$ is
missing transverse energy.
Therefore, we show in Fig.~\ref{BRsM1139}b a plot where these contributions are included (dashed red line). One can clearly see
that for $\mu \lsim 320~\gev$ the contributions of the invisible decays of the higher neutralinos (see Fig.~\ref{BRsM1139}a) are important. For comparison in
Fig.~\ref{BRsM1139}b the QFC branching ratios B$(\sg \to c \bar{c} \nt_1)$ and B$(\sg \to t \bar{t} \nt_1)$ are also shown.

The situation is quite different if $M_1 > M_2$. In this case, for $\mu \gsim M_2$ the $\nt_1$ is essentially a wino which does not couple to $\tilde{c}_R$ and $\tilde{t}_R$ leading to a small branching ratio of $\sg \to c \bar{t} (\bar{c} t) \nt_1$. On the other hand, a large branching ratio of $\sg \to c \bar{t} (\bar{c} t) \nt_2$ can be expected, because for $|\mu|  \gsim M_1$, $\nt_2$ becomes bino-like.

\begin{figure}
\centering
\subfigure[]{
   { \mbox{\hspace*{-0.6cm} \resizebox{8.4cm}{!}{\includegraphics{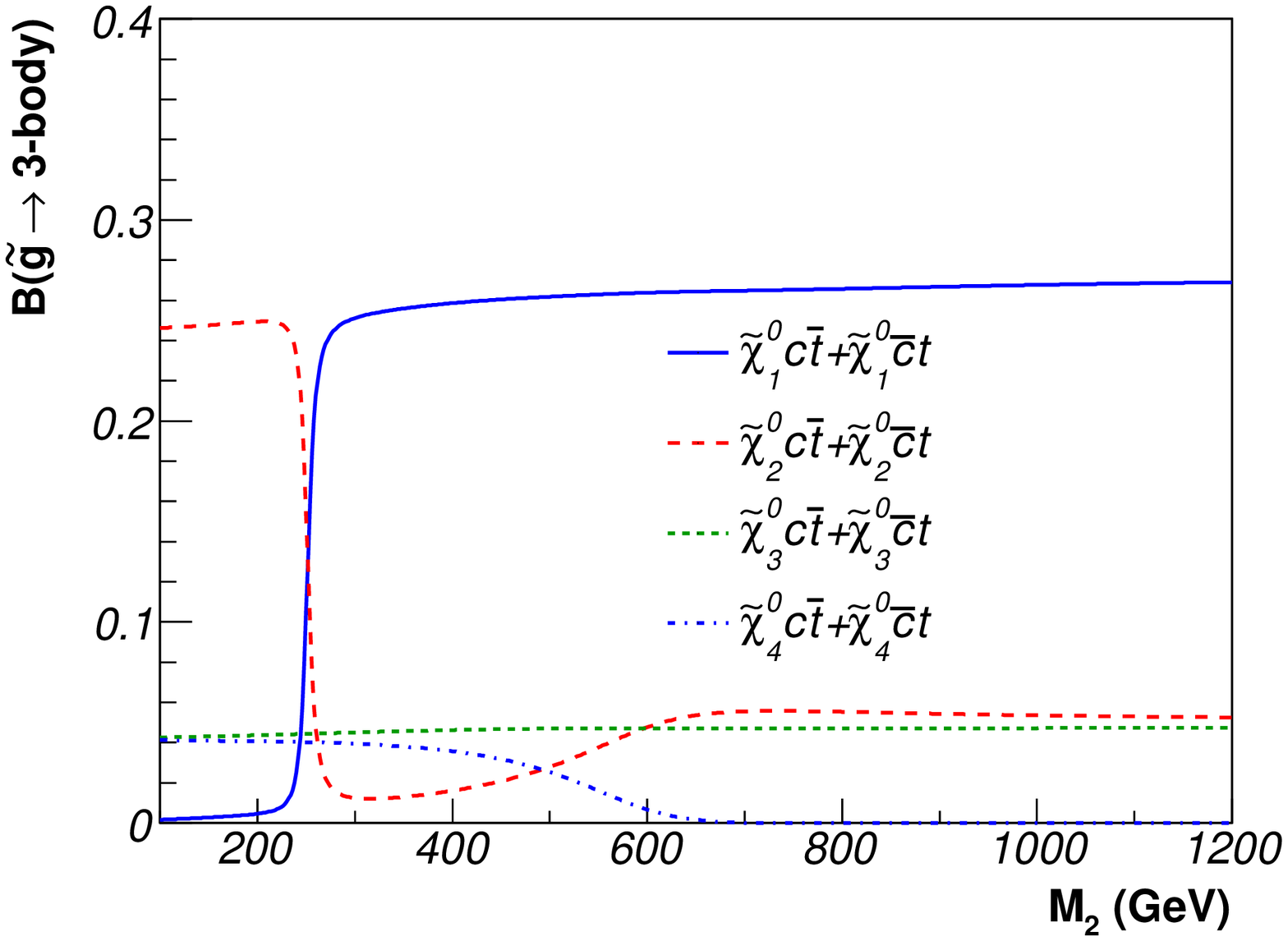}} \hspace*{-0.8cm}}}
   \label{M2depend1}}
 \subfigure[]{
   { \mbox{\hspace*{-0.3cm} \resizebox{8.4cm}{!}{\includegraphics{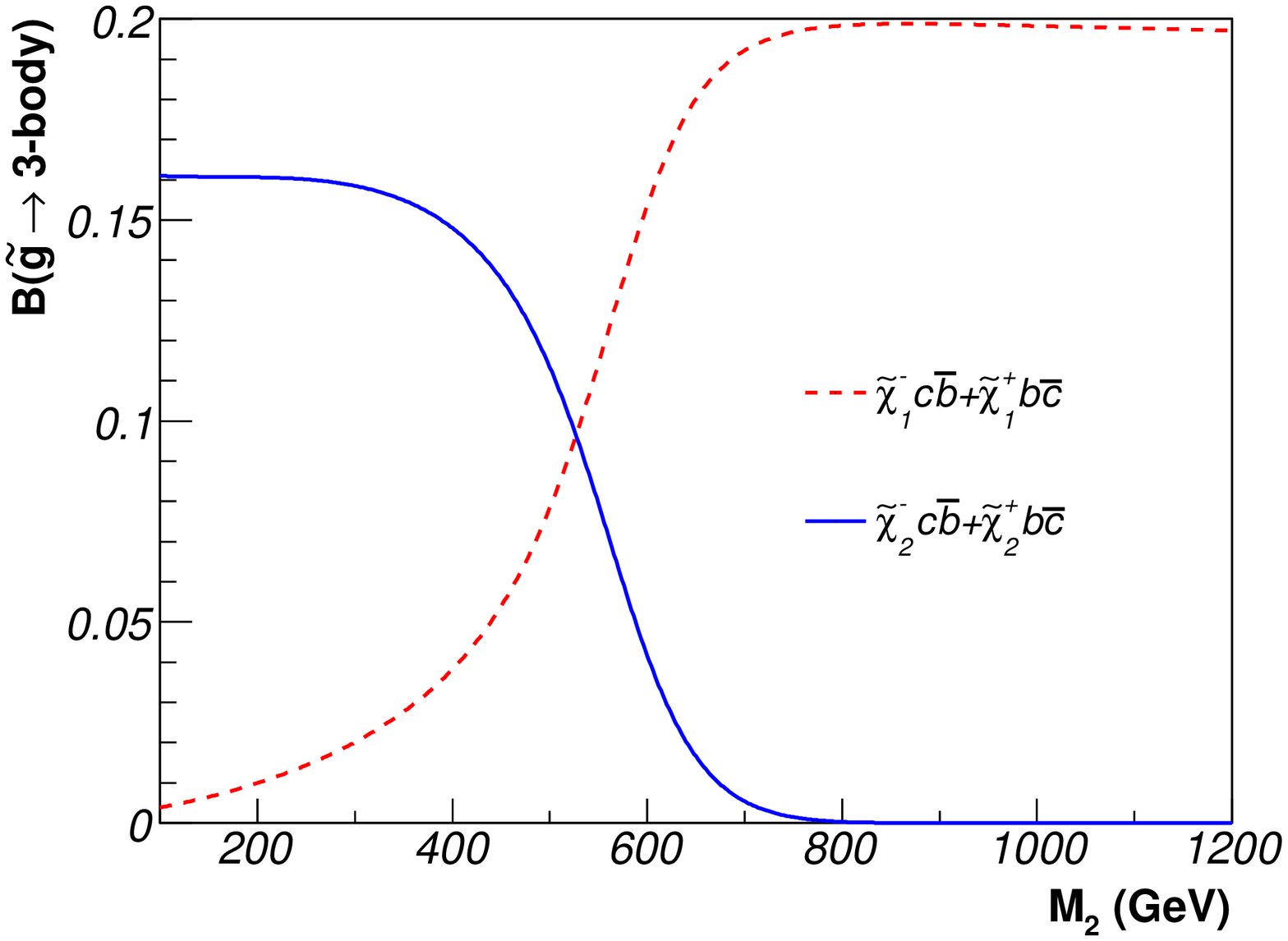}} \hspace*{-1cm}}}
  \label{M2depend2}}
\caption{The $M_2$ dependence of the QFV gluino decay branching ratios for $M_1=264~\gev,~\mu=600~\gev,~\delta^{uRR}_{23}=0.8$, the other QFV parameters being zero, and the
other parameters fixed as in Table~\ref{msg972soft}. (a) Branching ratios of the decays $\sg \to c \bar{t} \nt_i + \bar{c} t \nt_i$, $i=1,...,4$ as a function of $M_2$.
(b) Branching ratios of the decays $\sg \to c \bar{b} \ti{\chi}^-_k + \bar{c} b \ti{\chi}^+_k$, $k=1,2$ as a function of $M_2$.
\label{M2dep}}
\end{figure}

Next we discuss the dependence of the QFV gluino decay branching ratios on the $\rm{SU(2)} $ gaugino mass parameter $M_2$. In Fig.~\ref{M2dep} we show
this dependence fixing $M_1=264~\gev,~\mu=600~\gev,~\delta^{uRR}_{23}=0.8$, the other QFV parameters being zero, and the other parameters fixed as in Table~\ref{msg972soft}. In the shown range $m_{\sg}\approx 972~\gev$  and all the constraints mentioned in Section~\ref{sec:sq.matrix} are satisfied. As just explained above, in Fig.~\ref{M2dep}a, for $M_2 \lsim M_1$ the branching ratio of the decay  $\sg \to c \bar{t} (\bar{c} t) \nt_1$ is almost zero, while that for $\sg \to c \bar{t} (\bar{c} t) \nt_2$ is large ($\approx 24\%$).
For $M_2 \gsim M_1$ the
roles of $\nt_1$ and $\nt_2$ are interchanged and therefore the decay $\sg \to c \bar{t} (\bar{c} t) \nt_1$ becomes dominant with a branching ratio of about $25\%$. Note that the range $M_2 \lsim 100~\gev$ is
excluded by the LEP chargino mass limit mentioned in Section~\ref{sec:sq.matrix}: $m_{\ch_1} > 103~\gev$.
At $M_2 \approx \mu$ there is again a level crossing. The $\nt_4$ is higgsino-like for $M_2 \lsim \mu$ and becomes wino-like for $M_2 \gsim \mu$.

In Fig.~\ref{M2dep}b the branching ratios for the decays $\sg \to c \bar{b} (\bar{c} b) \ch_k$ as a function of $M_2$ are shown. The level crossing of $\ch_1$ and $\ch_2$ at $M_2 \approx \mu$ is clearly seen.

In the following we discuss in more detail a typical
scenario with a higgsino-like LSP ($\mu < M_1, M_2$) and one with a wino-like LSP ($M_2 < M_1, \mu$).
As an example for the higgsino-like LSP scenario, we choose the
parameters as given in Table \ref{higgsinoLSPparam1}, with the squark mass parameters as in Table \ref{msg972soft}. We fix the QFV parameter $\delta^{uRR}_{23}=0.8$ and the other QFV parameters equal zero. In this scenario all experimental and theoretical constraints mentioned in Section~\ref{sec:sq.matrix} are fulfilled. The relevant masses for the neutralinos and the charginos are given in Table \ref{higgsinoLSPparam2}. We show the most important QFV decay branching ratios in Table~\ref{higgsinoLSPBRs}. In this scenario $\nt_{1,2}$ and $\ch_1$ are almost higgsinos and
hence their couplings to $\ti{u}_1 (\sim \ti{c}_R+\ti{t}_R)$ are significantly enhanced by the large top-quark Yukawa coupling,
which results in the sizable branching ratios of the QFV decays into  $\nt_{1,2}$ and $\ch_1$.
Since B($\nt_2 \to \nt_1 \nu \bar{\nu}$)  is relatively large (= $18.4 \%$), the sum of the
branching ratios
for the $\sg$ decays into the final states ($c \bar{t}+ E_{\rm T}^{mis}$) and ($\bar{c} t+ E_{\rm T}^{mis}$)
is sizable ($= 9.0\%$).
The leptonic
$\ch_1$ decays $\ch_1 \to \mu^{\pm} \nu_\mu \nt_1$ and $\ch_1 \to e^\pm \nu_e \nt_1$ have a branching ratio of approximately
$11.2 \%$ each, because $W$ exchange dominates. Therefore, from the gluino decays into $\ch_1$ one gets final states
$b \bar{c} (\bar{b} c) +\mu^\pm (e^\pm) +E_{\rm T}^{mis}$ with a branching ratio of $5.0\%$. This has to be compared
with the expectation from MFV which is of the order of $10^{-4}$, as it is proportional to $|V_{cb}|^2$.
\begin{table}
\caption{
Weak scale parameters at  $Q=1$ TeV (except for $m_{A^0}$ being the pole mass),
the
corresponding neutralino and chargino masses and some important branching
ratios
for a scenario with a higgsino-like LSP, where $m_{\sg}=972~\gev$.}
\centering
\subtable[Weak scale parameters at $Q=1~{\rm TeV}$.]{
\centering
\small{
\begin{tabular}{|c|c|c|c|c|c|}
  \hline
 $M_1$ & $M_2$ & $M_3$ & $\mu$ & $\tan \beta$ & $m_{A^0}$ \\
 \hline \hline
 139~\gev  &  264~\gev &  800~\gev &  120~\gev & 10 &  800~\gev \\
  \hline
\end{tabular}}
\label{higgsinoLSPparam1}}
\subtable[Neutralino and chargino masses.]{
\centering
\small{
\begin{tabular}{|c|c|c|c|c|c|}
  \hline
  $\mnt{1}$ & $\mnt{2}$ & $\mnt{3}$ & $\mnt{4}$ & $\mch{1}$ & $\mch{2}$ \\
  \hline \hline
  $87.0~\gev$ & $133.1~\gev$ & $158.3~\gev$ & $310.1~\gev$ & $109.3~\gev$ & $310.1~\gev$ \\
  \hline
\end{tabular}}
\label{higgsinoLSPparam2}}
\subtable[Important branching ratios.]{
\centering
\small{
\begin{tabular}{|c|c|c|c|c|c|}
  \hline
 ${\rm B}(\sg \to c \bar{t} \nt_1$) & ${\rm B}(\sg \to c \bar{t} \nt_2$) & ${\rm B}(\sg \to b \bar{c} \tilde{\chi}^{+}_1)$ & ${\rm B}(\nt_2 \to \nt_1 \nu \bar{\nu})$ & ${\rm B}(\tilde{\chi}^{+}_1 \to \mu^+ \nu_\mu \nt_1)$ \\
 \hline \hline
 3.4 \% & 6.1 \% & 11.2 \% &  18.4 \% & 11.2 \% \\
  \hline
\end{tabular}}
\label{higgsinoLSPBRs}}
\end{table}
\begin{table}[h!]
\caption{
Weak scale parameters at $Q=1$ TeV (except for $m_{A^0}$ being the pole mass), the
corresponding neutralino and chargino masses and some important branching ratios for a scenario with a wino-like LSP, where $m_{\sg}=972~\gev$.}
\centering
\subtable[Weak scale parameters at $Q=1~{\rm TeV}$.]{
\centering
\small{
\begin{tabular}{|c|c|c|c|c|c|}
  \hline
 $M_1$ & $M_2$ & $M_3$ & $\mu$ & $\tan \beta$ & $m_{A^0}$ \\
 \hline \hline
 400~\gev  &  300~\gev &  800~\gev &  350~\gev & 10 &  800~\gev \\
  \hline
\end{tabular}}\label{AMSBparam1}}
\subtable[Neutralino and chargino masses.]{
\centering
\small{
\begin{tabular}{|c|c|c|c|c|c|}
  \hline
  $\mnt{1}$ & $\mnt{2}$ & $\mnt{3}$ & $\mnt{4}$ & $\mch{1}$ & $\mch{2}$ \\
  \hline \hline
  $275.5~\gev$ & $362.6~\gev$ & $376.4~\gev$ & $433.1~\gev$ & $280.0~\gev$ & $407.1~\gev$ \\
  \hline
\end{tabular}}
\label{AMSBparam2}}
\subtable[Important branching ratios.]{
\centering
\small{
\begin{tabular}{|c|c|c|c|c|c|}
  \hline
 ${\rm B}(\sg \to c \bar{t} \nt_1$) & ${\rm B}(\sg \to c \bar{t} \nt_2$) & ${\rm B}(\sg \to b \bar{c} \tilde{\chi}^{+}_1)$ & ${\rm B}(\nt_2 \to \nt_1 \nu \bar{\nu})$ & ${\rm B}(\tilde{\chi}^{+}_1 \to \mu^+ \nu_\mu \nt_1)$ \\
 \hline \hline
 2.5 \% & 6.3 \% & 5.8 \% &  4.1 \% & 13.2 \% \\
  \hline
\end{tabular}}
\label{AMSBBRs}}
\end{table}

Next we discuss a scenario where the LSP is wino-like, with the parameters as given in Table \ref{AMSBparam1}, the squark mass parameters as in Table \ref{msg972soft}, $\delta^{uRR}_{23}=0.8$ and the other QFV parameters being zero. Again, all the constraints are satisfied. The masses of the neutralinos and charginos are given in Table \ref{AMSBparam2}. The relevant QFV gluino decay branching ratios are shown in Table \ref{AMSBBRs}.
As in this case B($\nt_2 \to \nt_1 \nu \bar{\nu}$) = $4.1 \%$, the sum of the branchung ratios for the final states ($c \bar{t}+ E_{\rm T}^{mis}$) and ($\bar{c} t+ E_{\rm T}^{mis}$) is $6\%$.
As B($\tilde{\chi}^+_1 \to \mu^+ \nu_\mu \nt_1$) = B($\tilde{\chi}^+_1 \to e^+ \nu_e \nt_1$) = $13.2\%$, one has
B($\sg \to  b \bar{c} \mu^+ \nu_\mu \nt_1$) = B($\sg \to  \bar{b} c \mu^- \bar{\nu}_\mu \nt_1$) = B($\sg \to  b \bar{c} e^+ \nu_e \nt_1$)
=  B($\sg \to  \bar{b} c e^- \bar{\nu}_e \nt_1$) = $0.8\%$. Hence, the signature $b \bar{c}$ (or $\bar{b} c$) plus a lepton
($\mu^\pm$ or $e^\pm$) plus $E_{\rm T}^{mis}$ has a probability of about $3\%$.

Summarizing the discussion of this section we can say that the branching ratios of the QFV three-particle gluino decays depend not only on the generation mixing in the squark sector, but also quite strongly on the parameters of the neutralino/chargino sector.

\section{Measurability of the QFV gluino three-body decays}
\label{sec:measure}

We calculate the relevant gluino production cross sections at leading
order using
the WHIZARD/O'MEGA packages \cite{Whizard, Omega} where we have
implemented the model
described in Section~\ref{sec:sq.matrix} with squark generation mixing in its most general
form. We use
the CTEQ6L global parton density fit \cite{CTEQ6} for the parton distribution
functions and take $Q = m_{\ti{p}_1} + m_{\ti{p}_2}$ for the factorization scale,
where
$\ti{p}_1$ and $\ti{p}_2$ are the sparticle pair produced. The QCD coupling
$\alpha_s(Q)$ is also
evaluated (at the two-loop level) at this scale $Q$.
Due to the heavy squarks in our reference scenario of Table~\ref{msg972soft}, the
dominant gluino production
process at LHC is $pp \to \sg \sg X$, where $X$ contains beam-jets only. For the
scenario of Table~\ref{msg972soft} the
corresponding cross section is practically independent of $\d^{uRR}_{23}$ and is
about $170$~fb $(3$~fb) at
$\sqrt{s} = 14~{\rm TeV} ~(7~{\rm TeV})$. (Note that $m_{\sg} = 975~\gev$ and $972~\gev$ for
$\d^{uRR}_{23} = 0$ and 0.8,
respectively.) The sum of the cross sections of the other gluino production
processes, such as
$pp \to \sg \sq_1 X,~ \sg \sq_2 X,~ \sg \nt_1 X, ~ \sg \nt_2 X$, is two orders of
magnitude smaller than
that of $pp \to \sg \sg X$.
\begin{figure}
\begin{center}
\resizebox{11.5cm}{!}{\includegraphics{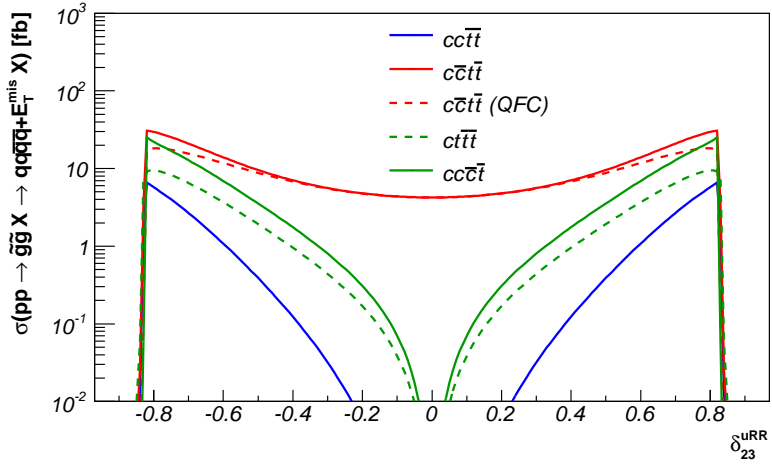}}
\end{center}
\caption{
Signal rates for $pp \to \sg \sg X$ at $\sqrt{s}=14~\rm{TeV} $ where at least one of
the gluinos
decays as $\sg \to c \bar{t} (\bar{c} t) \nt_1$, as a function of  $\delta^{uRR}_{23}$ with
the other QFV parameters
being zero and the other parameters fixed as in Table~\ref{msg972soft}. Shown are the
rates for the final
states with
$c c \bar{t} \bar{t} E_{\rm{T}}^{mis}$ (full blue line),  $c \bar{c} t \bar{t} E_{\rm{T}}^{mis}$ (QFV + QFC) (full red line),  $c \bar{c} t \bar{t} E_{\rm{T}}^{mis} $ (QFC only) (dashed red line),  $c t \bar{t} \bar{t} E_{\rm{T}}^{mis}$ (dashed green line),  $c c \bar{c} \bar{t} E_{\rm{T}}^{mis}$ (full green line).} \label{pptox}
\end{figure}

In Fig.~\ref{pptox} we show the signal rates due to $pp \to \sg \sg X$, with $X$ containing
beam-jets only, at
$\sqrt{s} = 14$~TeV, where at least one of the pair-produced gluinos decays as
$\sg \to c \bar{t} (\bar{c} t) \nt_1$, as a function of $\delta^{uRR}_{23}$ for the scenario of Table~\ref{msg972soft}. All the
constraints mentioned in
Section~\ref{sec:sq.matrix} are satisfied in the range $|\d^{uRR}_{23}| \lsim 0.85$.
The rate of the final state $c c \bar{t} \bar{t} E_{\rm{T}}^{mis} X$, produced in the case
when both gluinos
decay like $\sg \to c \bar{t} \nt_1$,
reaches 7~fb for $|\d^{uRR}_{23}| = 0.8$ (full blue line), yielding 700 events for an integrated luminosity of $100~\fb$.
The charge conjugated final state $\bar{c} \bar{c} t t E_{\rm{T}}^{mis} X$ has the same rate.
The full red line shows the rate for the QFV case  where one gluino decays as $\sg \to c \bar{t} \nt_1$ and the other one as $\sg \to \bar{c} t \nt_1$ plus the QFC
case with one $\sg \to c \bar{c} \nt_1$ and the other $\sg \to t \bar{t} \nt_1$.
The dashed red line presents the rate for the QFC case only.
One can see
that the QFV signal has a rate of about $14$~fb for $|\delta^{uRR}_{23}| = 0.8$.
The green lines show the case of one gluino decaying as
$\sg \to c \bar{t} \nt_1$, and the other one as $\sg \to c \bar{c} \nt_1$ (full green line) or $\sg \to t \bar{t} \nt_1$ (dashed green line). These rates
reach $25$~fb and $10$~fb, respectively, for $|\delta^{uRR}_{23}| = 0.8$. The charge conjugated final states have the same rates.
%
\begin{figure}
{\setlength{\unitlength}{1mm}
\begin{center}
\hspace*{-15mm}
\begin{picture}(130,50)
\put(30,0){ \mbox{\resizebox{8.5cm}{!}{\includegraphics{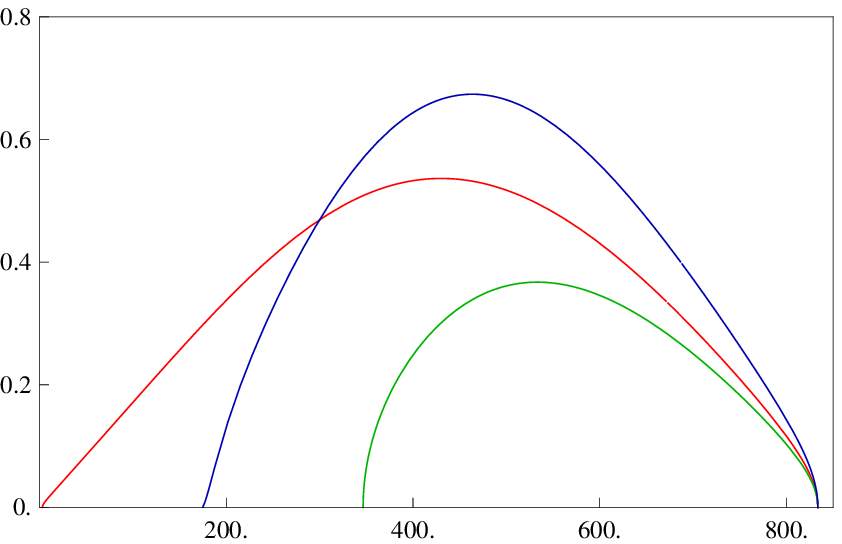}}}}
\put(96,36.5){\mbox{$c \bar{t} + \bar{c} t$}}
\put(62,-7.5){ \mbox{ $M_{u_j \bar{u}_k} [\gev$]}}
\put(74,15.5){\mbox{$t \bar{t}$}}
\put(50.5,26){\mbox{$c \bar{c}$}}
\hspace*{2cm}
\begin{sideways}\small{
\put(4,37){\mbox{ $10^3\times \frac{{\rm d\Gamma}(\sg \to u_j \bar{u}_k \nt_1)}{\Gamma_{tot} (\sg) ~ dM_{u_j \bar{u}_k}} ~[\rm{GeV}^{-1}]$}}
}
\end{sideways}
\end{picture}
\end{center}}
\vspace*{1cm}
\caption{Invariant mass distributions of two up-type quarks from the decay $\sg \to u_j \bar{u}_k \nt_1$, with $\delta^{uRR}_{23}=0.8$, the other QFV parameters being zero, and the other parameters fixed as in Table~\ref{msg972soft}.}
\label{inv_mass}
\end{figure}

A characteristic
feature of a three-particle decay of the gluino is the invariant mass distribution of the two produced quarks.
We calculate the invariant mass distributions ${\rm d\Gamma} (\sg \to u_j \bar{u}_k \nt_1)/(\Gamma_{tot} (\sg) ~ dM_{u_j \bar{u}_k})$, where $M_{u_j \bar{u}_k}$ is the invariant mass of the two-quark system $u_j \bar{u}_k$, $M_{u_j \bar{u}_k}^2 = (p_{u_j}+p_{\bar{u}_k})^2$.
In Fig.~\ref{inv_mass} we show these distributions for $ \sg \to t \bar{t} \nt_1, c \bar{c} \nt_1, c \bar{t} \nt_1+\bar{c} t \nt_1$ with $\delta^{uRR}_{23}=0.8$ and the other QFV parameters being zero for the scenario of Table~\ref{msg972soft}.
In contrast to the case where the squarks are lighter than the gluino and the gluino decays via real squarks \cite{Ba2}, no edge structure appears. However, the thresholds and the shapes of the distributions are very different. The endpoint is at
$(m_{\sg} -m_{\nt_{1}})$. The thresholds are at $2 m_c, m_c + m_t$ and $2 m_t$, respectively. Measuring these distributions could be helpful
for separating the QFV decays into $ c \bar{t} \nt_1+\bar{c} t \nt_1$ from the QFC decays.

A typical $\sg \sg$ production
event has at least four large-$p_T$ jets and large $E_{\rm{T}}^{mis}$. An event with a QFV gluino decay $\sg \to c \bar{t} \nt_1 (\bar{c} t \nt_1)$ should contain at least one top (anti-top) quark in the final state, which must be identified. This is possible by using the decay $t \to b W^\pm$ with the $W^\pm$ decaying into two jets.
For this purpose a special method was proposed in \cite{N}. Charm tagging would be extremely helpful for a
clear identification of the QFV gluino decay  $\sg \to \bar{c} t \nt_1 (c \bar{t} \nt_1)$.
If this is not possible one could search for the decay $\sg \to \bar{q} t \nt_1 (q \bar{t} \nt_1), q \ne t$. Typical signatures of QFV $\sg$ pair events are:
$t ( {\rm or}~\bar{t})+ {\rm 3~jets} + E_{\rm{T}}^{mis} + X$, $t + t~
({\rm or}~\bar{t} + \bar{t}) +{\rm  2~jets} + E_{\rm{T}}^{mis} + X$ and $t +
t + \bar{t} ~({\rm or}~\bar{t} + \bar{t}+ t) +{\rm  1~jet} +
 E_{\rm{T}}^{mis} + X$ ,
where $X$ contains beam-jets only. Note, that the signal events $t +
t~
({\rm or}~\bar{t}+\bar{t}) + {\rm 2 jets} +E_{\rm{T}}^{mis} + X$ can practically
not be
produced in the MSSM (nor in the SM) with QFC.

The rate of a possible SUSY background from pair production of squarks,
such as $p p \to \sq \bar{\sq} X$ with $\sq \to c \nt_1$, $\bar{\sq} \to \bar{t} \nt_1$ is much
smaller
than that of the signal of gluino pair production due to the larger squark
masses.
As shown in the SUSY searches by ATLAS and CMS~\cite{ATLAS, CMS}, the SM
backgrounds, such as
QCD multijets, $W^\pm$ + jets,   $Z^0$ + jets, $t \bar{t}$ and single top
production, can be strongly reduced by appropriate selection cuts.

\section{Conclusions}

We have studied QFV decays of gluino within the MSSM in the case that all squarks are heavier than the gluino and,
hence, the gluino has only three-particle decays. Starting from the most general squark mass matrix, we have assumed mixing
between the second and the third squark generations in the up and down sectors.
We have taken into account
all relevant experimental constraints from SUSY particles and Higgs searches as well as from precision data in the B meson sector.
Furthermore we have respected the vacuum stability conditions for the trilinear coupling matrices.
It has turned out that of all QFV parameters the parameters $\delta^{uRR}_{23},~\delta^{dRR}_{23}$ and $\delta^{LL}_{23}$
play the most important role in our study. We have concentrated on the QFV decays $\sg \to c \bar{t} (\bar{c} t ) \nt_1$
and  $\sg \to b \bar{s} (\bar{b} s) \nt_1$ which presumably have the clearest signatures for the presence of QFV in the MSSM.
We have studied these within a prototype scenario with gluino mass $m_{\sg} \approx 1$ TeV and squarks with masses between $\sim 1$ TeV and $\sim 3$ TeV,
where the lightest squarks are $\ti{c}_R-\ti{t}_R$ mixtures (or $\ti{s}_R-\ti{b}_R$ mixtures).
These QFV decay branching ratios can reach up to 40\%~(35$\%$).
We have paid special attention to the dependence of the QFV decay branching ratios on the chargino/neutralino parameters. In this
context we have considered three cases where the lightest neutralino is bino-, wino- and higgsino-like, respectively.
We have found that the QFV decay branching ratios depend strongly also on the chargino/neutralino parameters.

As in our scenario the squarks are heavier than the gluino, the dominant gluino production process at LHC is gluino pair production,
$pp \to \sg \sg X$. We have calculated the rates for the various signatures stemming from QFV gluino decays as well as the
invariant mass distributions of the two-quark system $c \bar{t}+\bar{c} t, c \bar{c}, t \bar{t}$ in the final states.
We have found that the rates of the resulting QFV signatures, like $pp \to t c \bar{c} \bar{c} E_{\rm T}^{mis} X$ and $pp \to t t \bar{c} \bar{c} E_{\rm T}^{mis} X$,
are significant at LHC.
This could have an important influence on the search for gluinos
and the determination of the basic MSSM  parameters at LHC.

\section*{Acknowledgments}

This work is supported by the "Fonds zur F\"orderung der
wissenschaftlichen Forschung (FWF)" of Austria, project
No. I 297-N16,
and by the DFG, project No.
PO-1337/2-1.
B. H. acknowledges support by the Landes-Exzellenzinitiative
Hamburg.


\begin{appendix}

\end{appendix}


\end{document}